\documentclass[fleqn,usenatbib]{mnras}
\usepackage{newtxtext,newtxmath}

\usepackage[T1]{fontenc}
\usepackage{ae,aecompl}

\usepackage{graphicx}	
\usepackage{amsmath}	
\usepackage{amssymb}	
\usepackage{breqn}		
\usepackage{mathtools}
\usepackage{multirow}
\usepackage{tabularx}
\usepackage{hyperref}    
\usepackage{etoolbox}
\makeatletter
\patchcmd\@combinedblfloats{\box\@outputbox}{\unvbox\@outputbox}{}{%
  \errmessage{\noexpand\@combinedblfloats could not be patched}%
}%
\makeatother

\newcommand{\mhaloe}{M_{h}}

\newcommand{\msune}{M_{\odot}}

\title{A Statistical Semi-Empirical Model: Satellite galaxies in Groups and Clusters}

\author[P. J. Grylls et al.]{
Philip J. Grylls,$^{1}$\thanks{E-mail: P.Grylls@soton.ac.uk}
F. Shankar,$^{1}$\thanks{E-mail: F.Shankar@soton.ac.uk}
L. Zanisi,$^{1}$
M. Bernardi$^{2}$
\\
$^{1}$Department of Physics and Astronomy, University of Southampton, Highfield, SO171BJ, UK\\
$^{2}$Department of Physics and Astronomy, University of Pennsylvania, PA 19104, USA\\
}

\date{Accepted XXX. Received YYY; in original form ZZZ}

\pubyear{2018}

\begin{document}

\label{firstpage}
\pagerange{\pageref{firstpage}--\pageref{lastpage}}
\maketitle

\begin{abstract}
We present STEEL a STatistical sEmi-Empirical modeL designed to probe the distribution of satellite galaxies in groups and clusters. Our fast statistical methodology relies on tracing the abundances of central and satellite haloes via their mass functions at all cosmic epochs with virtually no limitation on cosmic volume and mass resolution. From mean halo accretion histories and subhalo mass functions the satellite mass function is progressively built in time via abundance matching techniques constrained by number densities of centrals in the local Universe. By enforcing dynamical merging timescales as predicted by high-resolution N-body simulations, we obtain satellite distributions as a function of stellar mass and halo mass consistent with current data. We show that stellar stripping, star formation, and quenching play all a secondary role in setting the number densities of massive satellites above $M_*\gtrsim 3\times 10^{10}\, M_{\odot}$. We further show that observed star formation rates used in our empirical model over predict low-mass satellites below $M_*\lesssim 3\times 10^{10}\, M_{\odot}$, whereas, star formation rates derived from a continuity equation approach yield the correct abundances similar to previous results for centrals.

\end{abstract}




\section{Introduction}
Cold dark matter predicts a hierarchical assembly of dark matter haloes, which grow in mass through accretion from smaller units to increasingly larger systems. The complex mass assembly of haloes is recorded in the dark matter merger trees. After infall into larger systems, accreted satellite haloes will eventually merge with the parent halo through stripping and orbital decay from dynamical friction. Dark matter structure formation has been studied in numerous numerical works with  \citep[e.g.,][]{Lemson2006HaloCosmogony,Klypin2016} and without \citep[e.g.,][]{Vogelsberger2014IntroducingUniverse,Fattahi2016TheSelection} the baryonic component. Dark matter-only simulations have been performed on very large scales, with cosmological volumes up to several $Gpc^3$ \citep{Potter2017PKDGRAV3:Surveys}. In comparison, due to the computational complexities introduced by hydrodynamics, simulations with baryons have been limited to orders of magnitude smaller volumes, around a hundred $Mpc^3$. The discrepancy in simulated volumes inevitably leads to structures with intrinsically lower number densities limiting their usefulness in studying less common but important galaxy populations, such as massive satellite galaxies residing in relatively small parent haloes. Hydrodynamical simulations have resolution and time constraints imposed by the computational power they can leverage, ultimately being forced to turn to sub-grid analytic physical recipes below the particle or mesh resolution. 

Semi-analytic models \citep[e.g.,][]{Granato2004AHosts,Baugh2006AApproach,Monaco2007TheNuclei,Hirschmann2012GalaxySimulations,Shankar2013SizeUniverse} make comprehensive predictions on the evolutionary patterns of galaxies by following the growth and assembly along dark matter merger trees. Semi-analytic models are necessarily characterized by a number of input assumptions and free parameters associated with different formation mechanisms. Semi-analytic models make the admirable effort towards a holistic view of galaxy formation, but they are known to suffer from degeneracies in assumptions and related parameters \citep[e.g.][]{Lapi2011DarkModels,Gonzalez2011EVOLUTION4}, and could still be limited by volume effects.

In the past decades much attention has been devoted to probing the connection between dark matter and galaxies in a more empirical fashion, attempting to build a self-consistent picture of dark matter/galaxy assembly and distribution making use of more observationally-driven methods. Halo Occupation Distribution models were introduced as a first step in this direction \citep[e.g.,][]{COORAY2002halostructure, Berlind2002TheMass}. These models are designed to predict the average number of central and satellite galaxies in haloes by simultaneously matching observed number densities and clustering at a given epoch. Halo occupation models by construction do not account for the full assembly history of the galaxies and may be susceptible to assembly bias effects \citep[e.g.,][]{Zentner2014GalaxyRelationship, Dalal2008HaloFormation}.

Abundance matching between the galaxy luminosity/stellar mass function and halo mass function is a semi-empirical technique that has been widely adopted in the last two decades to constrain, on a statistical basis, the mean relation between stellar mass and host halo\footnote{Here and throughout this work ``host halo'' refers to the dark matter (sub)halo associated with a given galaxy residing at the (sub)halo's center.} mass \citep[e.g.,][]{Kravtsov2004TheDistribution, Vale2004LinkingLuminosity,Yang2004PopulatingSurveys,Shankar2006NewFormation}. 
In its simplest form, abundance matching relies on the basic idea that larger galaxies reside in larger dark matter haloes, and relative abundances are used to create a (mean) stellar mass-halo mass (SMHM) mapping $M_* = SMHM(M_h, z)$ at any relevant epoch \citep{Yang2003ConstrainingGalaxies, Moster2010, Behroozi2010A4}. This technique has been used as a flexible tool to predict the full mass and structural assembly history of galaxies along dark matter merger trees in a full cosmological context. 

Semi-empirical models \citep[e.g.,][]{Hopkins2009HOWMERGERS,Cattaneo2011HowMass,Zavala2012,Shankar2014} were conceived as models where galaxies are assigned to haloes at any epoch along dark matter merger trees via abundance matching relations, thus allowing to track the full stellar mass evolution in, for example, the main progenitor branch of the merger tree. Semi-empirical models, although unavoidably more restrained in scope, avoid the often heavy parameterizations incorporated in semi-analytic models as galaxy stellar masses and other properties (e.g., disc sizes) are empirically assigned to dark matter haloes. In a semi-empirical approach a minimum number of free parameters are included in the models allowing exploration of specific aspects of galaxy formation minimizing the danger of degeneracies. Despite their flexibility, traditional semi-empirical models can be subject to volume limitation effects due to the finite number of merger trees and/or dark matter haloes. Both semi-analytic and semi-empirical models can therefore suffer from low number statistics especially at high stellar masses when built on top of dark matter simulations.

In this work we present STEEL a STastical sEmi-Empirical modeL built using a novel, fast and flexible methodology to probe the number densities and average properties of galaxies at any given stellar mass, epoch and environment, with negligible computational limitation on volume or mass resolution. This is particularly useful when aiming at simultaneously probing, as in this work, the distribution of massive centrals as well as massive satellites in lower mass haloes. The method employed in this paper starts from tracking backwards in time the mean accretion histories of small bins in halo mass, weighted by the halo mass function. At each timestep, satellite galaxy number densities are associated to the main halo mass bin via the corresponding time growth in subhalo mass function. The main aim of this work is to predict satellite number densities as a function of stellar mass and halo mass. We will show that the dominant parameter controlling the abundance of massive satellites is the dynamical friction timescale. Nevertheless, the evolution of satellite galaxies, as central ones, is of course a result of a combination of star formation, quenching, and stellar stripping. We will probe the (limited) impact on our results of each one of these processes. 

In what follows we adopt the Planck cosmology with $(\Omega_m, \Omega_{\Lambda}, \Omega_{b}, h, n, \sigma_8) = (0.31, 0.69, 0.05, 0.68, 0.97, 0.82)$\footnote{We note that Planck's best-fit cosmology has slightly different parameters than those adopted in some of the observational probes included in this work, such as the stellar mass functions ($(\Omega_m, h = (0.30, 0.70)$). However, we have checked that correcting the stellar mass function volumes and luminosities to the same Planck's cosmology yields essentially indistinguishable results in the implied stellar mass-halo mass relation. We will thus ignore such tiny differences in cosmological parameters in what follows. We choose to retain the halo cosmology as the full dark matter backbone of the model presented in this work is built and calibrated against the Planck Cosmology \citep{PlanckCollaboration2015PlanckParameters}.}. We use $M_h$ and $M_*$ to differentiate between halo and stellar mass respectively, we name $M_{h, cent}$ and $M_{h, sat}$ central and satellite respectively (we direct readers to Appendix \ref{app:TechAcro} for a complete glossary of the acronyms used throughout this work). Halo masses are defined as virial masses, unless stated otherwise. Wherever relevant, we always assume a \cite{Chabrier2003GalacticFunction} stellar initial mass function.

\section{Data}
\label{sec:Data}
In this work we take as a reference the Sloan Digital Sky Survey Data Release 7 (SDSS-DR7) from \cite{Meert2015ASystematics,Meert2016ABands}, with improved galaxy photometry. In brief, the data is originally from the SDSS DR7 spectroscopic sample \citep{Abazajian2009THESURVEY} containing $\sim 670,000$ galaxies fitted with a S\'ersic + exponential model \citep[PyMorph;][and references therein]{Meert2015ASystematics}, with associated halo masses \citep{Yang2012EVOLUTIONHALOS} updated to the new photometry. The improved photometry provides a number of advantages. For instance, the two-dimensional fit more accurately captures the high mass end of the stellar mass function \citep{Bernardi2013TheProfile,Bernardi2017TheProfile,Bernardi2017ComparingLight} and uses the mass-to-light ratio from \citet{Mendel2014ASURVEY}.

At higher redshifts, ($0.3 < z <3.3$) stellar mass functions (SMF) are from the COSMOS2015 catalogue \citep{Davidzon2017TheSnapshots}. In this survey masses are derived from multiwavelength spectral energy distribution fitting, including ultra-deep and infrared photometry. We note \citet{Davidzon2017TheSnapshots} use \citet{Bruzual2003Stellar2003} stellar population synthesis models which are consistent within $\lesssim 0.15$ dex with the mass to light ratios adopted in the SDSS catalogue. Both data sets assume a \cite{Chabrier2003GalacticFunction} stellar initial mass function.

\section{Methods: A new ``statistical'' approach to semi-empirical modelling}
\label{sec:Method}

Semi-analytic models continue to provide invaluable constraints on the formation and evolution of galaxies. They admirably attempt to analytically model from zero order the full evolution of galaxies since the Big Bang to the local Universe \citep[e.g.,][]{Menci2006TheModels}.
This inevitably requires specific input assumptions and related parameters on, for example, the rate of star formation at different epochs and/or environments.

Semi-empirical models are instead characterized by a ``bottom-up'' approach, where the least possible assumptions and associated parameters are initially included in the models. Gradually, additional degrees of complexities can be included in the model, wherever needed. Semi-empirical models in this respect are a very powerful complementary tool to semi-analytic models or even cosmological hydrodynamic simulations. Their goal is in fact to provide unique constraints to only specific aspects of galaxy evolution, such as predicting, as in this work, the distribution of satellites in a given dark matter parent halo\footnote{Here and throughout this work ``parent halo'' refers to the central halo containing subhaloes/satellite galaxies.}.

Semi-empirical models \citep{Hopkins2009HOWMERGERS,Cattaneo2011HowMass,Zavala2012,Shankar2014} are broadly characterized by the following steps: 
\begin{enumerate}
\item Central galaxies, associated to the main progenitor branch\footnote{The main progenitor is defined in traditional merger trees as the largest progenitor halo from any given halo merger, usually the largest halo at any time step. In this work shown in Panel B of Figure \ref{fig:StochasticTree} we simply have the average growth of this halo but it is convenient to visualize in the traditional sense.} of the host dark merger tree, are initially defined as gas-rich with disc like profiles at early epochs, as supported by deep HST observations in CANDELS (e.g., Huertas-Company et al. 2015, 2016). All their stellar mass and structural properties are assigned via empirical, time-dependent relations, more significantly the stellar mass-halo mass relation \cite[e.g.][]{Moster2013}.
\item Central galaxies are then re-initialized at each time step during the evolution and can gradually transform their morphology via ``in-situ'' processes, such as (more or less) violent disc instabilities, and/or ``ex-situ'' processes, such as mergers. 
\item Satellite galaxies are those associated to each dark matter branch merging to the main progenitor. They are assigned all the mass and structural properties of a central galaxy in a typical central halo of the same mass at the time of infall. Satellites may further evolve in stellar mass given their specific star formation rate at infall, possibly accompanied by some stellar stripping.
\end{enumerate}

The typical semi-empirical model described above is a powerful and very competitive tool to explore trends in the expected mass and/or structural evolution of galaxies in a given interval of stellar mass for a given set of parameters in, e.g., amount of stripping and/or orbital energy \citep{Shankar2015}. Based, by design, on closely following the evolution of each dark matter merger tree, the basic semi-empirical approach discussed above is still limited by the resolution and volume of the background dark matter simulation. This could be particularly restraining and computationally expensive especially when modelling very massive galaxies and their accompanying large volumes. 

In this paper we propose a statistically-based semi-empirical approach. We explore a large range of stellar masses, with virtually no limitation in sample size, and still retain the high degree of flexibility characteristic of a semi-empirical model. The core steps in building the dark matter backbone of STEEL are summarized as follows (see diagram in Figure \ref{fig:StochasticTree}): 
\begin{enumerate}
\item At the redshift of interest $\bar{z}$ we start from the halo mass function to compute the abundances of (parent/central) haloes (panel A). 
\item Each parent/central halo mass is then followed backwards in time following its average mass growth history, $<M_{\rm halo}(\bar{z})>$, as expected from extended Press-Schechter \citep{Press1974} and/or numerical simulations (panel B).
\item At each time step we then calculate the difference in subhalo population
between $z$ and $z+dz$ to estimate the expected average number density and masses of subhaloes accreted onto the main progenitor in the redshift interval $dz$ (panel D). 
\end{enumerate}

We will show how using analytic halo mass functions (HMF) and a ``statistical accretion history'', in place of halo catalogues and merger trees, we are able to simultaneously examine massive central haloes, and massive satellites in lower mass parents. The latter are usually characterized by low number densities and can be omitted in more traditional approaches which rely on cosmological volumes. For example, massive satellite galaxies with stellar mass $M_*>10^{11} M_{\odot}$ residing in relatively smaller parent haloes of mass $M_h\sim 10^{12.75}\, M_{\odot}$, have predicted number densities of $\sim 10^{-6}\, Mpc^{-3}$ (See Figure \ref{fig:Distrbution_Tdyn}, Panel 3), which would correspond to $\sim 1$ galaxy in a volume of $\sim 100$ $[Mpc/h]^3$, a typical volume adopted in state-of-the-art cosmological hydrodynamical simulations. On the other hand, even in larger dark matter-only simulations such as the Bolshoi ($250$ $[Mpc/h]^3$) \citep{Klypin2016}, one would expect at the most $\sim 16$ of these galaxies. As massive satellite are so rare in other simulations, a robust statistical determination of abundances is therefore vital to fully constrain the connection between galaxy formation and hierarchal assembly predicted by $\Lambda CDM$ cosmology at all mass ranges \citep{Neistein2013A2011}.

In this section we detail each step of this statistically-based semi-empirical approach. We start by outlining in Section \ref{sec:SDMAH} the analytic model of dark matter accretion histories characterizing the mean growth of the main progenitor halo and accreted subhaloes including subhalo merging times. We then discuss in Section \ref{sec:Abn} how we populate dark matter haloes with galaxies, using the stellar mass-halo mass relation for central galaxies and its evolution with redshift. Finally, in Section \ref{sec:SatEvo}, we discuss the evolutionary processes that affect the satellite galaxies whilst they reside in their dark matter parent haloes.

\subsection{Statistical Dark Matter Accretion History}
\label{sec:SDMAH}

\begin{figure*}
	\centering
	\includegraphics[width = 0.95\linewidth]{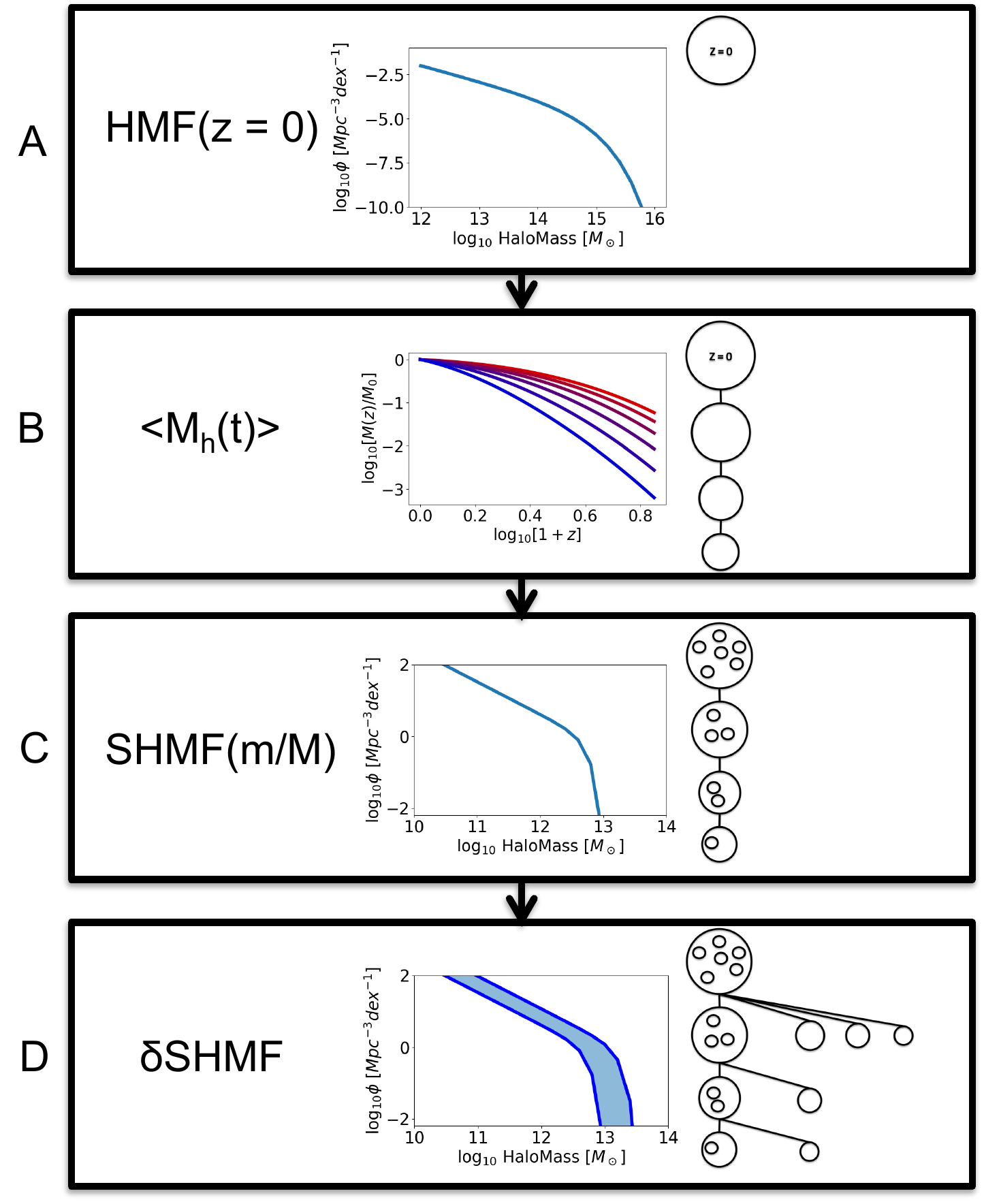}
	\caption{We show the main steps in building the statistical dark matter backbone of STEEL described in \ref{sec:SDMAH}. Each panel shows the feature from a traditional merger tree and the statistical function used to replace it. A: The HMF is used to calculate the number densities of central haloes. B: Average mass growth histories calculate the size of each mass bin at previous epochs. C: The SHMF is used to populate each central at each redshift with subhaloes. D: The average number densities of accreted subhaloes at each epoch, are calculated by comparing the SHMF over consecutive redshift steps.}
	\label{fig:StochasticTree}
\end{figure*}

The core principle of our methodology is to treat parent haloes, and satellites galaxies/haloes, as ``average'' populations avoiding issues with volume and resolution as described previously. Specifically, the goal of this paper is to reproduce the observed distribution of $z = 0.1$ satellite galaxies with mass  $M_* > 10^{9} M_{\odot}$. In this subsection we detail step by step construction of the backbone of STEEL complemented with a graphic representation in Figure \ref{fig:StochasticTree}.

\subsubsection{Central Haloes}

We start by considering a fine grid of central dark matter haloes ranging from $\mhaloe=10^{11}\, \msune$ to $\mhaloe=10^{15}\, \msune$ at redshift $z=0$. Their number densities are given by the halo mass function (HMF), which is obtained using HMFcalc \citep{Murray2013HMFcalcFunctions}. HMFcalc includes a number of different mass functions. For this work we use the HMF from \cite{Tinker2010THETESTS}\footnote{We adopt the COLOSSUS Python package \citep{Diemer2017COLOSSUS:Halos} for all halo mass conversions required.}. The halo mass function provides the number densities of haloes in a given mass bin (Figure \ref{fig:StochasticTree}, Panel A).
The average mass growth histories of all main progenitors with mass in the bin of halo mass $[\mhaloe,\mhaloe+d\mhaloe]$, are then calculated using the analytic model from \cite{Bosch2014ComingWells}\footnote{This model further improves on the seminal work by \cite{Parkinson2008GeneratingTrees}, which was aimed at reproducing numerical merger trees, optimized with small redshift steps minimizing the development of systematic errors at late cosmic epochs.}. This provides the average ``main progenitor'' branch of a traditional merger tree for each mass bin at $z = 0$ (Figure \ref{fig:StochasticTree}, Panel B.)

\subsubsection{Assigning Subhaloes to Parent Haloes}

In order to predict the number of satellite galaxies, we must associate to each parent/central halo the number and mass of subhaloes they are expected to contain. To achieve this we use the subhalo mass function (SHMF). The SHMF describes the expected distribution of subhaloes, of mass $M_{h,sat}$, in a given parent halo of mass $M_{h,cent}$, as a function of $M_{h,sat}/M_{h,cent}$. Multiple definitions for the SHMF exist depending on the way a subhalo is defined. In this work we use two definitions of the SHMF. The first is the unevolved SHMF (USHMF), which describes the total subhaloes accreted over a parent halo's lifetime. In the unevolved SHMF any merging or stripping in the subhaloes occurring after infall are ignored. A number of groups have been able to constrain the unevolved SHMF \citep{Giocoli2008AnalyticalHaloes,Jiang2016StatisticsFunctions}. In what follows, we use the latest rendition of the unevolved SHMF by \citet{Jiang2016StatisticsFunctions}, which is calibrated against the Bolshoi simulation\footnote{ We direct the interested reader to \citet{Jiang2016StatisticsFunctions} for further discussion of the unevolved SHMF as well as of other SHMFs, such as the evolved SHMF where the number densities are effected by both subhalo stripping and mergers.}.

\begin{figure}
	\centering
	\includegraphics[width = \linewidth]{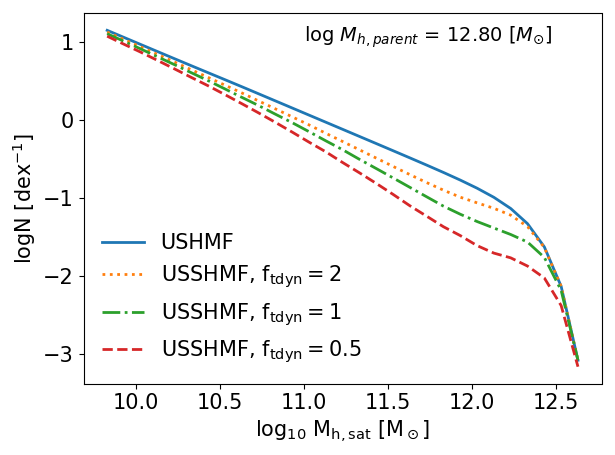}
	\caption{Comparison between the unevolved SHMF (solid line) and three unevolved surviving SHMF  (dotted lines) for a parent halo of mass $\log$ $M_{h,parent}$ $[M_{\odot}] = 12.80$. The factor $f_{t_{dyn}}$ is applied to the merging timescales of the haloes. Lower factors correspond to lower unevolved surviving SHMF where more sub haloes have merged}
	\label{fig:SHMF_clus}
\end{figure}

The second definition we use in this work is the unevolved ``surviving'' SHMF (USSHMF). Subhalo masses are assumed frozen at infall but the subhalo number densities can reduce compared to the unevolved SHMF as the unevolved surviving SHMF accounts for subhalo disappearance due to tidal disruption in the parent halo. We show in Figure \ref{fig:SHMF_clus} for a representative parent halo of mass $\log M_{h,cent} M_{\odot} = 12.80$, the unevolved SHMF and three unevolved surviving SHMF characterized by different dynamical friction timescales $\tau_{dyn}$, as described in Section \ref{sec:Timescale}. Larger $\tau_{dyn}$ lead to a milder reduction in subhalo number densities as subhaloes take longer to merge with the parent halo. Lower $\tau_{dyn}$ are less effective in reducing the number densities of smaller subhaloes which are more likely to have dynamical friction timescales comparable to or larger than the Hubble time at $z = 0$.

\subsubsection{Average Subhalo Accretion}
At each redshift step along the mass growth histories we calculate the unevolved SHMF associated to the parent halo mass. What we create is equivalent to the substructure found in traditional merger trees (Figure \ref{fig:StochasticTree}, Panel C). However, unlike traditional methods, our statistical approach is able to probe ``rare'' subhaloes without running prohibitively large volumes of merger trees.

For each timestep we can now calculate a mass function describing the number density of subhaloes accreted onto the population of central haloes in the halo mass bin $[M_{h,cent}(z),$ $M_{h,cent}(z) + dM_{h,cent}(z)]$. The latter is achieved by differentiating the unevolved SHMF across two neighbouring redshift steps $z$ and $z+dz$, we can calculate the average number density of subhaloes of any given mass $M_{h, sat}$ that are accreted in the redshift interval $dz$ onto the main progenitor haloes with mass in the bin $[M_{h,cent}(z),$ $M_{h,cent}(z) + dM_{h,cent}(z)]$,
\begin{equation}
\label{eqn:deltSHMF}
\begin{split}
&\delta USHMF[z, M_{h,cent},M_{h,sat}] =  \\
&USHMF\Big(\frac{M_{h,sat}}{M_{h,cent}(z)}\Big) - USHMF\Big(\frac{M_{h,sat}}{M_{h,cent}(z + \delta z)}\Big)
\end{split}
\end{equation}
In this way the unevolved subhalo accretion history ($\delta USHMF$) is retrieved for all main progenitor haloes at all redshifts.

\begin{table*}
\centering
\caption{Parameters used in the model and how they are constrained.}
\label{my-label}
\begin{tabularx}{\linewidth}{cXc}
\hline
Parameter                          & Description                                                                                                                                                  & Equation\\ \hline
      							   & Free Parameters
								   &\\ \hline
								   $\mathrm{f_{tdyn}}$                         & A factor we include to test the impact of shortening or lengthening dynamical times of all satellites. This is a free parameter we use to find the best fit to satellite distributions.
								   & \ref{eqn:tdyn}\\    \hline 							   
                                   & Constrained Parameters
                                   &\\ \hline
                                   A,B,C,D                            & Dynamical friction fitting parameters given by \cite{McCavana2012TheMergers}                                                                                & \ref{eqn:Tmerge}\\
                                   $\mathrm{M_n, N}$, $\beta$, $\gamma$        & Abundance Matching parameters, constrained by the central SMF and HMF at z = 0                                                                                                                                                  & \ref{eqn:MosAbn}\\
                                   $\mathrm{M_{n,z}, N_z}$, $\beta_z$, $\gamma_z$        
                                   & Abundance Matching evolution parameters, constrained using SMF and the HMF at z >0.                                                                                                                                      & \ref{eqn:MosAbn}\\

s, M, $\alpha$                     & Star formation parameters, constrained by observation given in \cite{Tomczak2016THE4}. We also give values to fit our continuity derived star formation rate in Section \ref{sec:SFR}                                                                                                                         & \ref{eqn:SFR}\\
$\tau_q$, $\tau_f$, $\mathrm{M_{cutoff}}$ & Parameters describing the quenching time of satellites. These parameters are all adapted from models in the literature \citep{Wetzel2013GalaxyUniverse,Fillingham2016UnderStripping}                                                                                                 & \ref{eqn:Cutoff} - \ref{eqn:SFR_Quench}\\                                                            &               \\
$\eta_{strip}$                     & Stripping fraction from \cite{Cattaneo2011HowMass}                                                                                                                                                                                                                                    & \ref{eqn:Strip}
\end{tabularx}
\end{table*}

\subsubsection{Subhalo Merging Timescale}
\label{sec:Timescale}
From the unevolved subhalo accretion history we need to isolate which subhalo mass bins survive at each epoch. The sum of all the surviving subhaloes (at each epoch) then yields the unevolved surviving SHMF. A key parameter used to calculate the unevolved surviving SHMF is the ``observability timescale''(or survival time) of each  subhalo mass bin $[M_{h,sat}(z),$ $M_{h,sat}(z) + dM_{h,cent}(z)]$ associated to a parent halo mass bin $[M_{h,cent}(z),$ $M_{h,cent}(z) + dM_{h,cent}(z)]$. This timescale is equivalent to the merger timescale $\tau_{merge}$ of a subhalo of mass $M_{h,sat}$ in a parent halo mass $M_{h,cent}$. To calculate $\tau_{merge}$ we use the routines in Equation \ref{eqn:Tmerge}, derived from N-body simulations \citep{Boylan-Kolchin2008}

\begin{equation}
\label{eqn:Tmerge}
\begin{split}
\tau_{merge} =& \\
(f_{t_{dyn}}\tau_{dyn})& \frac{A(M_{h, cent}/M_{h,sat})^B}{\ln(1+M_{h, cent}/M_{h, sat})} \exp \Big(C\frac{J}{J_c(E)}\Big) \Big( \frac{r_c(E)}{r_{vir}} \Big)^D,
\end{split}
\end{equation}

where A=0.9, B=1.0, C=0.6, D=0.1 \citep{McCavana2012TheMergers}. The factor $\tau_{dyn}$ is given by \citep{Jiang2016StatisticsFunctions},

\begin{equation}
\label{eqn:tdyn}
\tau_{dyn} = 1.628 h^{-1} \mathrm{Gyr} \Big(\frac{\Delta_{vir}(z)}{178}\Big)^{-\frac{1}{2}} \Big(\frac{H(z)}{H_0}\Big)^{-1} \, .
\end{equation}

Our method of considering average halo mass and accretion histories does not allow tracking single orbits and associated orbital energies. We assume instead an average orbit circularity of 0.5 \citep{Khochfar2006OrbitalHalos}, thus reducing the dependence on the angular momentum and radial components, $\frac{J}{J_c(E)}$ and $\frac{r_c(E)}{r_{vir}}$, to a constant. In other words, this approximation is consistent with the approach of taking the average expected orbits of subhaloes at fixed parent halo mass.
The key parameter of our analysis is the factor $f_{t_{dyn}}$ included in Equation \ref{eqn:Tmerge}.
The fudge factor $f_{t_{dyn}}$ takes into account the systematic uncertainties induced by numerical resolution effects in N-body simulations which are unable to resolve the full merging timescales of subhaloes and/or the satellite galaxies they host \citep{Bosch2018DisruptionFiction}. The parameter $f_{t_{dyn}}$ increases or decreases the dynamical times of  ``merging'' satellites enabling an exploration of the effect of dynamical time on the final number density distributions of satellite galaxies at any given epoch.

\subsubsection{Surviving Subhalo Population}
At each redshift we can now use the unevolved subhalo accretion history and the observability timescale $\tau_{merge}$ to calculate the total observable subhalo population associated with any given parent halo mass bin $[M_{h,cent}(z),$ $M_{h,cent}(z) + dM_{h,cent}(z)]$, i.e. the unevolved surviving SHMF (shown by the dashed lines in Figure \ref{fig:SHMF_clus}). To compute the implied total number densities of unmerged subhaloes with mass $[M_{h,sat}(z),$ $M_{h,sat}(z) + dM_{h,sat}(z)]$ at any redshift of interest we convolve the unevolved surviving SHMF with the HMF,

\begin{equation}
\label{eqn:GSHMF}
\begin{split}
N(M_{h, sat}, z) =&\\
\int USSHMF&\Bigg(\frac{M_{h, sat}}{M_{h, cent}}\Bigg)HMF(M_{h, cent}, z)dM_{h, cent}.
\end{split}
\end{equation}

\begin{figure}
	\centering
	\includegraphics[width = \linewidth]{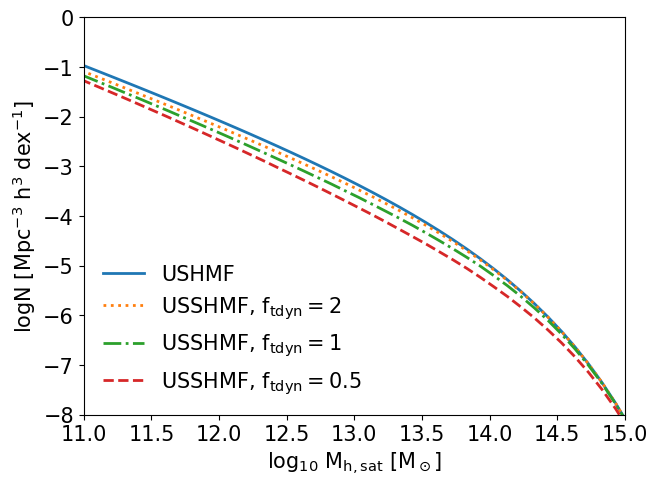}
	\caption{Example of the `total' unevolved SHMF (solid line) and three `total' unevolved surviving SHMF (dotted lines) corresponding to three different $f_{t_{dyn}}$ factors.}
	\label{fig:SHMF}
\end{figure}

Figure \ref{fig:SHMF} shows the total observable subhalo population for different  $f_{t_{dyn}}$ similar to the SHMF in Figure \ref{fig:SHMF_clus}. Furthermore, via appropriate abundance matching algorithms detailed below, we can assign corresponding satellite galaxies to the unevolved subhalo accretion history and obtain the distribution of satellites in a given parent halo mass bin $[M_{h,cent}(z),$ $M_{h,cent}(z) + dM_{h,cent}(z)]$ by assuming the satellites follow the same merging timescales as their host subhaloes. Finally, we convolve the latter distributions of satellite galaxies with the HMF to create the total observed satellite stellar mass function (SSMF). Both the distribution of satellite galaxies and the satellite SMF can then be used to constrain $f_{t_{dyn}}$ as demonstrated in Section \ref{sec:NDandDist}.
\subsection{Abundance Matching: Building the local populations of central and satellite galaxies}
\label{sec:Abn}

We now describe how we populate dark matter haloes with galaxies using the abundance matching technique. To compute the distribution of satellites in parent haloes we populate subhaloes at infall, following the central stellar mass - halo mass relation (SMHM). For the latter, we adopt a parametric form similar to that proposed by \cite{Moster2010},
\begin{equation}
\label{eqn:MosAbn}
\begin{split}
M_*(M_h, z) &= 2M_hN(z)\Big[ \Big( \frac{M_h}{M_{n}(z)}\Big) ^{- \beta(z)} + \Big( \frac{M_h}{M_{n}(z)}\Big)^{\gamma(z)} \Big ]^{-1}\\
N(z) &= N_0.1 +N_z\Big(\frac{z-0.1}{z+1}\Big)\\
M_{n}(z) &= M_{n,0.1} +M_{n,z}\Big(\frac{z-0.1}{z+1}\Big)\\
\beta(z) &= \beta_0.1 +\beta_z\Big(\frac{z-0.1}{z+1}\Big)\\
\gamma(z) &= \gamma_0.1 +\gamma_z\Big(\frac{z-0.1}{z+1}\Big).
\end{split}
\end{equation}

\subsubsection{Local universe $z = 0.1$}

To constrain the $z = 0.1$ parameters ($M_{n}$,  $N$, $\beta$, $\gamma$, $\sigma$), we construct a SMF from the HMF and the SMHM relation with initial parameters close to previous results \citep{Moster2013,Buchan2016,Shankar2014}. The HMF provides the number density for each bin of parent halo mass in the range $[M_{h,cent}(z),$ $M_{h,cent}(z) + dM_{h,cent}(z)]$. Using the SMHM relation, each bin of parent halo mass is associated to a Gaussian distribution of stellar masses with width controlled by the scatter parameter ($\sigma$). Multiplying by the halo mass number density we convert this distribution into galaxy number densities, which are then added to the relevant stellar mass bins of the SMF in construction.  This operation is repeated over each bin of the HMF. The resulting SMF is then compared to the SDSS central SMF. We vary the parameters on a grid, and minimize root mean square between the constructed SMF and the data. In our abundance matching framework, we specifically distinguish between central and satellite stellar mass functions by taking advantage of the SDSS galaxy-halo catalogue described in Section \ref{sec:Data}.

For comparison with previous works we apply the same method as described above to the total galaxy/halo population using the unevolved surviving SHMF ($f_{t_{dyn}} = 1.0$), the HMF, and the total SMF. Our total result uses single-epoch \footnote{A multi-epoch approach considers the infall time of subhaloes, and routines are used for stripping, star formation, and quenching of satellite galaxies. The full set of parameters are fit simultaneously \citep[e.g.][]{Moster2013}.\label{MultiEp}} abundance matching: we apply the SMHM relation to the total HMF at $z=0.1$ with no consideration of the infall time of subhaloes.

\begin{figure*}
	\centering
	\includegraphics[width = \linewidth]{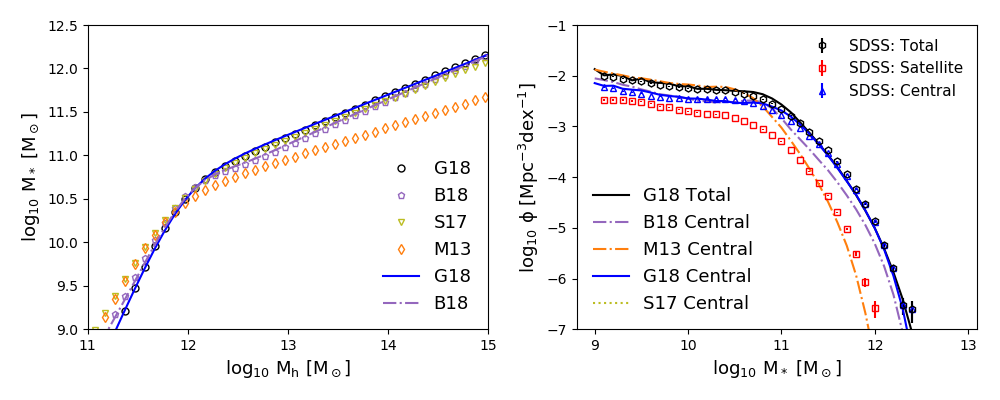}
	\caption{Left Panel: The SMHM relation generated from this work (G18, black circles), \citet[][B18]{Behroozi2018UniverseMachine:Z=0-10}, \citet[][S17]{Shankar2017RevisitingRedshift} and \citet[][M13]{Moster2013}. Relations derived from the total SMF/HMF are shown with open symbols, while relations derived with the central SMF/HMF are show as solid lines. Right Panel: Central SMF generated from the models using a central only fit, we also show the total SMF from this work (solid black line) from a single-epoch fit. Open symbols show the SDSS central (squares), satellite (triangles) and Total (hexagons) stellar mass functions.}
	\label{fig:SMF_Abn}
\end{figure*}

Table \ref{tab:Abn} reports the best-fit parameters using the method above, for central galaxies and total galaxy population. In the left panel of Figure \ref{fig:SMF_Abn} we show with a solid blue line the $z=0.1$ stellar mass function for SDSS central galaxies only, abundance-matched with the central HMF (no subhaloes). For comparison we also include the abundance matching fits by \cite{Moster2013}, \cite{Shankar2017RevisitingMass} and \citet[][using the ``true'' stellar masses, in table J1]{Behroozi2018UniverseMachine:Z=0-10}. The \textit{total} stellar mass function is reproduced by \cite{Moster2013} and \cite{Shankar2017RevisitingMass} without distinguishing between centrals and satellites galaxies in the SMF. We find our final results for the high mass slope to be steeper than the \cite{Moster2013} but very similar to the ones by \cite{Shankar2017RevisitingMass}, as expected given that the latter have tuned their fitting routines to the same galaxy stellar mass function as the one adopted in this paper. At the low mass end our SMHM relation tends to appear slightly steeper than previous works. At least with respect to \cite{Shankar2017RevisitingMass}, this is mainly caused by the fact that our results are based on the updated surviving subhalo mass function by \cite{Jiang2016StatisticsFunctions} which is somewhat steeper than the subhalo mass correction suggested by \cite{Behroozi2013THE0-8} and adopted by \cite{Shankar2017RevisitingMass}. Our low mass slope is, as expected, fully consistent with \cite{Behroozi2018UniverseMachine:Z=0-10} as their subhaloes have been extracted from the Bolshoi-Planck simulation \citep{Klypin2016}. The right panel of Figure \ref{fig:SMF_Abn} shows the total, central, and satellite stellar mass functions as extracted from our SDSS catalogues (black circles, blue triangles and red squares, respectively) compared with the central SMF generated using the central HMF and the relations from \cite{Moster2013}, \cite{Behroozi2018UniverseMachine:Z=0-10}, and our best-fit relations. The solid black line shows our ``total'' relationship where the full population of central and satellite galaxies are retrieved with single-epoch abundance matching.

From this comparison we conclude that for the SMF and HMF used in this work single epoch abundance matching is acceptable as there is little evolution at late times. However, if the HMF or SMF were to be strongly evolving at low redshift, this approach would yield a result inconsistent with matching the central population. We further argue firmer constraints are set using the central SMHM relation where there is higher confidence in the populations being matched.

\begin{table}
\centering
\caption{Parameters for Equation \ref{eqn:MosAbn}. The $z=0.1$ parameters are constrained to 2 d.p. or 2 s.f., whichever is smaller. The $z>0.1$ parameters are constrained to 1 d.p. or 1 s.f., whichever is smaller. }
\label{tab:Abn}
\begin{tabular}{cccccc} \hline
                                                                     & $M_{n}$     & N      & $\beta$ & $\gamma$ & $\sigma$ \\ \hline
\begin{tabular}[c]{@{}c@{}}Central\\ z=0.1\end{tabular}                 & 11.95 & 0.032  & 1.61  & 0.54  & 0.11     \\
\begin{tabular}[c]{@{}c@{}}Total\\ z=0.1\end{tabular}                 & 11.89 & 0.031  & 1.77  & 0.52  & 0.10     \\ \hline
\begin{tabular}[c]{@{}c@{}}Evolution\\ z \textgreater 0.1\end{tabular} & 0.4   & -0.02 & -0.6 & -0.1 & N/A  
\end{tabular}
\end{table}

\subsubsection{Redshift evolution $z > 0.1$}

We assume that at any redshift satellite galaxies at infall follow the same scaling relations as centrals of equal parent halo mass but then may evolve independently afterwards. Therefore we require a SMHM relation calibrated at all redshifts to initialize satellites in subhaloes at infall.

Keeping fixed the $z = 0.1$ parameters, the redshift-dependent parameters in Equation \ref{eqn:MosAbn} are fit with a similar iterative method. A complication arises as the high redshift data do not distinguish central and satellite SMF. 
We therefore use a multi-epoch\textsuperscript{\ref{MultiEp}} approach without satellite evolution to fit the evolution parameters.
We continue to generate central SMF as above then add a satellite SMF and compare to the total SMF from the data minimizing the sum of the root mean squares over eight redshift steps. We generate the satellite SMF\footnote{Under the assumption that satellites are ``frozen'' in stellar mass after infall. We will discuss the impact of relaxing this assumption in Section \ref{sec:SatEvo}} from the model for each set of parameters in our parameter space. In Appendix \ref{app:Abn} we show the result of our iterative method applied to the \cite{Davidzon2017TheSnapshots} and SDSS stellar mass functions.

\begin{figure*}
	\centering
	\includegraphics[width = \linewidth]{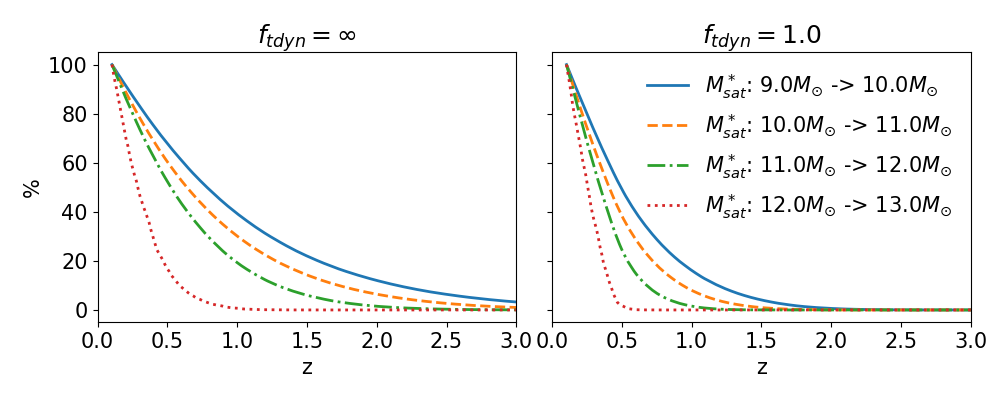}
	\caption{We here show the percentage of satellites observed at a $z = 0.1$ as a function of their redshift of accretion $z > 0$. It can be seen that massive satellites observed at $z = 0.1$ are accreted more recently than smaller satellites. At $z = 0.5$ less that 50\% of the total satellites observed at $z = 0.1$ have been accreted, and at $z = 0.1$ this falls to less than 20\%. }
	\label{fig:AccretionTime}
\end{figure*}

It is important to realize at this stage that, as demonstrated in Figure \ref{fig:AccretionTime}, the vast majority (>60-80\%) of $z = 0.1$ satellites are accreted below $z<1.0$ irrespective of the exact dynamical timescale adopted\footnote{We note this does not imply an absence of satellite galaxies at higher redshift, only that these satellites at higher redshift mostly do not survive to be observed at $z = 0.1$.}. In this redshift range the evolution of the SMHM relation has been demonstrated to be rather weak \citep[e.g.,][and references thereof]{Moster2013,Behroozi2013THE0-8,Shankar2014,Bernardi2016TheEvolution}. Thus, we expect that our results are robust against possible variations in the redshift dependence of the SMHM relation used to initialize infalling satellites, as demonstrated below.

We will now focus on specifically reproducing the local \textit{satellite} stellar mass function (red squares in the right panel of Figure \ref{fig:SMF_Abn}).
\subsection{Satellite evolution after infall}
\label{sec:SatEvo}

After infall, satellites are expected to evolve their stellar mass and structure through several processes such as star formation, stellar stripping, and quenching. In this Section we briefly describe each of these processes and the way we have modelled them in this work. We will then discuss in Section \ref{subsec:EvoPro} how these processes may impact, if at all, the predicted local abundances of satellite galaxies. 

\subsubsection{Star Formation Rates}
\label{sec:SFR}

We use the star formation rate (SFR) parameterization from \cite{Lee2015A1.3} with parameters\footnote{These parameters are derived by fitting data from ZFORGE in combination with far-IR imaging from \textit{Spitzer} and \textit{Herschel} in the range $0.5<z<4$. In this work we extrapolate their fits down to $z = 0$, as this is consistent with the SFR measured by  \cite{Salim2007UVUniverse} at lower redshifts.} from \cite{Tomczak2016THE4} where $s_0$ and $M_0$ have units $log(M_{\odot})$ and $M_{\odot}$ respectively,

\begin{equation}
\begin{split}
\label{eqn:SFR}
\log[\psi(z, M_*)] &= s_0(z) - \log \Big[ 1 + \Big(\frac{M_*}{M_0(z)}\Big)^{-\alpha(z)}\Big] \\
s_0(z) &= 0.195 + 1.157z - 0.143(z^2) \\
\log[M_0(z)] &= 9.244 + 0.753z - 0.090(z^2) \\
\alpha(z) &= 1.118.
\end{split}
\end{equation}
As discussed below, we will also need to explore different recipes for SFRs. To this purpose, following a continuity-equation approach \citep{Leja2015RECONCILINGFUNCTION}, we have refitted the parameters in Equation \ref{eqn:SFR} to match the SFR distributions expected from self-consistently growing the satellite stellar mass function in our models (for further details on this see Appendix \ref{app:SFRCont}). The fit to the resulting SFRs is given by

\begin{equation}
\begin{split}
\label{eqn:SFR_CE}
\log(\psi(z, M_*)) &= s_0(z) - \log \Big[ 1 + \Big(\frac{M_*}{M_0(z)}\Big)^{-\alpha(z)}\Big] \\
s_0(z) &= 0.6 + 1.22z - 0.2(z^2) \\
\log(M_0(z)) &= 10.3 + 0.753z - 0.15(z^2) \\
\alpha(z) &= 1.3 - 0.1z.
\end{split}
\end{equation}
In all cases the SFR is included in our models with a log-normal scatter of 0.3 dex \citep{Leja2015RECONCILINGFUNCTION}.

\subsubsection{Quenching}
\label{sec:QuenchingModels}

\begin{figure}
	\centering
	\includegraphics[width = \linewidth]{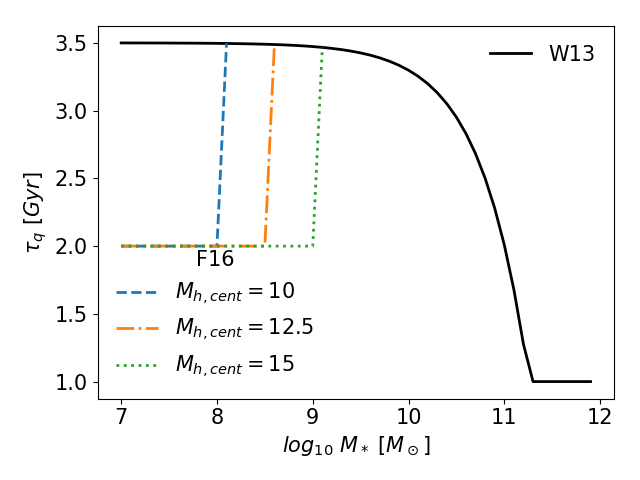}
	\caption{The solid line shows the \citet[][W13]{Wetzel2013GalaxyUniverse} model for quenching. The dashed lines show the host halo dependent reduction in quenching time from \citet[][F16]{Fillingham2016UnderStripping} for three example host masses $\log10$ $M_{h, cent} = 10, 12.5, 15$ as labelled. Larger hosts are able to reduce the quenching time of larger satellites}.
	\label{fig:QuenchFig}
\end{figure}

We also include the ability to quench star formation in satellite galaxies after infall. It has been suggested that satellites undergo a ``delayed-then-rapid'' quenching \citep{Wetzel2013GalaxyUniverse}. The latter model envisions that satellites continue to form stars at the same rate as central galaxies of comparable stellar mass for a time $\tau_q$ after infall, and then quench rapidly over a timescale $\tau_{f}$. This quenching is proposed for satellites with stellar mass above $M_*\gtrsim 10^9 M_{\odot}$, with a minimum $\tau_q$ $\sim$ $1$ $Gyr$. For galaxies below $M_* \lesssim 10^{9} M_{\odot}$, we however adopt the more recent results by \citet{Fillingham2016UnderStripping}, who put forward a parent halo mass dependent cutoff,
\begin{equation}
\label{eqn:Cutoff}
log(M_{cutoff}) = 9 log(M_{\odot}) - (15 log(M_{\odot}) - log(M_{h,host}))/5log(M_{\odot}) ,
\end{equation}
below which satellite galaxies all share the same quenching time $\tau_q=2 Gyr$.
The rapid quenching timescale $\tau_{f}$ can be expressed as \citep{Wetzel2013GalaxyUniverse},
\begin{equation}
\label{eqn:tauf}
\tau_f = -0.5 log(M_{*,sat}) + 5.7Gyr .
\end{equation}
We set a minimum $\tau_{f}$ of $0.2$ $Gyr$ for all galaxies. This rapid quenching begins at times $t > \tau_q$ after infall, and it is approximated by an exponential decay which is longer for larger satellites, as can be inferred from Equation \ref{eqn:SFR_Quench}. The lookback time at which a galaxy begins fast quenching is then $t_q = t_{infall} - \tau_q$. After this time the satellite no longer follows the SFR of a typical central galaxy. Figure \ref{fig:QuenchFig} illustrates the quenching model where the dashed/coloured lines demonstrate the halo mass dependence in cutoff mass.

The SFR during the satellite infall is then given by
\begin{equation}
\label{eqn:SFR_Quench}
SFR(t, M_*) = SFR(t, M_*)
\begin{dcases}
\psi(z(t), M_*), & \text{} t > t_q \\
\psi(z(t_q), M_*)e^{\big[-\frac{t_q-t}{\tau_f}\big]}. & \text{} t < t_q
\end{dcases}
\end{equation}
If at any point a satellite galaxy has a SSFR below $10^{-12}$ $M_{\odot}$ $yr^{-1}$, it is assumed to be fully quenched and assigned a SSFR of $10^{-12}$ $M_{\odot}$ $yr^{-1}$, plus a log-normal scatter of 0.3 dex.

\subsubsection{Mass Recycling}
\label{sec:MassRecycling}
During infall some of the new stellar mass formed may be returned to the interstellar medium after stellar death. To incorporate this mass recycling process we use the fractional mass loss as parameterized in \cite{Moster2018Emerge10}
\begin{equation}
\label{eqn:f_ml}
f(\tau_{ml}) = 0.05 \ln \Big(\frac{\tau_{ml}}{1.4 Myr}+1\Big) \, .
\end{equation}
It follows that the mass-loss rate (MLR) at a given time is dependent on the star formation history, and it is calculated as
\begin{equation}
\label{eqn:MLR}
MLR(t) = \frac{ \sum_{t' = t_{infall}}^{t} SFH(t')(f[t' - (t-\delta t)]-f[t' - t]) }{\delta t} .
\end{equation}
In Equation \ref{eqn:MLR} the starformation history (SFH) is the mass of stars formed in each previous time step. The actual stellar mass growth $\dot{M}_{*, sat}$ is then given by the difference between the star formation rate and the mass loss rate (MLR)
\begin{equation}
\label{eqn:Mdot}
\dot{M}_{*, sat}(t) = SFR(M_{*, sat}, t) - MLR(t) .
\end{equation}

\subsubsection{Stripping}
\label{sec:Stripping}
The stellar stripping is implemented following the empirically-based formalism suggested by \citet{Cattaneo2011HowMass}. Satellite galaxies strip stellar mass proportionally to the ratio of the host subhalo and the parent halo
\begin{equation}
\label{eqn:Strip_Coff}
M_{*,sat} = M_{*,sat}(1-\eta_{strip})^{\tau_{strip}},
\end{equation}
where $\tau_{strip}$ is the dynamical friction timescale in orbital time. For our model, where the orbital circularity is assumed to take the average value, this becomes the dynamical friction time \citep{Chandrasekhar1943DYNAMICALFRICTION} times a constant given by \cite{Jiang2008AClustering} using the average orbital circularity. Following \citet{Cattaneo2011HowMass} the latter can be written,
\begin{equation}
\label{eqn:Strip}
\tau_{strip} = \frac{1.428}{2\pi}\frac{M_{h,cent}/M_{h, sat}}{\ln(1+M_{h,cent}/M_{h,sat})}. 
\end{equation}

We set as a reference $\eta_{strip}$ = 0.4 as suggested by \citet{Cattaneo2011HowMass} who showed that larger values would be inconsistent with the observational constraints in the local universe. If at the time of observation, say $z=0$, a galaxy's full dynamical time is not yet passed, a time-dependent reduction factor is applied to the amount of total stellar stripping given by Equation \ref{eqn:Strip}.

\section{Number Densities and Distributions of massive satellites}
\label{sec:NDandDist}

Section \ref{sec:Method} has laid out our methodology, which is essentially based on progressively building up the (surviving) satellite number densities over cosmic time based on the input dynamical merger timescales. Our first goal is now to predict the local ($z=0.1$) satellite SMF and the distribution of massive satellites as a function of parent halo mass, which we will compare with our refined SDSS galaxy-halo catalogue introduced in Section \ref{sec:Data}. When computing  the distribution of satellites as a function of parent halo mass we will show both full number densities, as well as fractional distributions to better highlight the ``skewness'' of the predicted distributions with respect to the data. The latter will be simply computed as
\begin{equation}
\label{eqn:FracPlot}
F(dM_h) = \frac{N(>10^x)|_{dM_h}}{N(x)},
\end{equation} 
where $N(>10^{x})$ is the total number (density) of satellites above a threshold stellar mass $x= \log M_{*}$, and the $N(>10^{x})|_{dM_h}$ is the number of these that reside within the halo mass bin $[M_h, M_h+dM_h]$.

\begin{figure*}
	\centering
	\includegraphics[width = \linewidth]{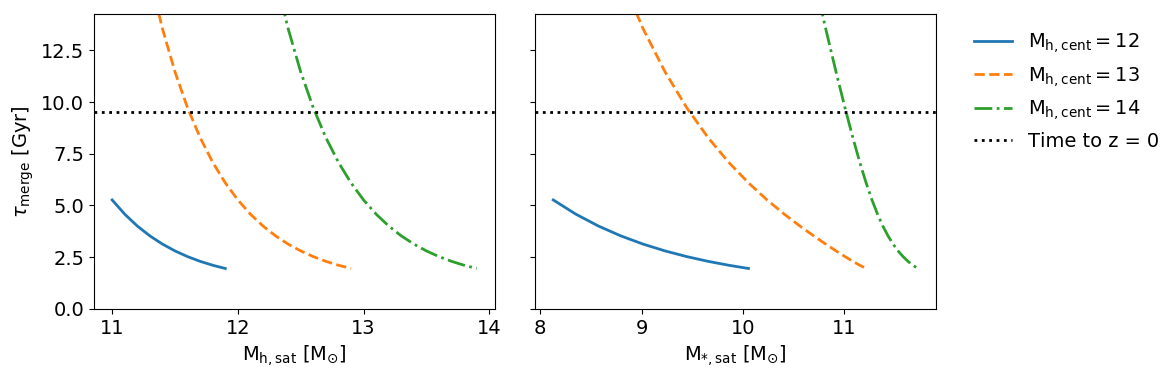}
	\caption{Range of merging timescales for a range of subhalo (left) and satellite (right) masses when accreted onto three different host masses: $\log$ $M_{h, cen}$ $[M_{\odot}]$ = 12 (blue solid), 13 (orange dashed) and 14 (green dot-dashed). The dotted black line shows the time to redshift $z = 0$, i.e. the minimum amount of time a satellite would need to survive to be observable in the local universe.}
	\label{fig:Tdyn_M}
\end{figure*}

The results discussed in the next Sections rely heavily on the concept of mass ratio affecting the merging timescales of subhaloes and/or satellite galaxies. To ease the discussion of our findings, we provide Figure \ref{fig:Tdyn_M} as visual guide to how merging times depend on subhalo/satellite masses in three different parent halo masses. In the left hand panel we plot dynamical time against subhalo mass for a subhalo accreted at $z =  1.5$. The black dotted line is the time (in Gyr) to $z = 0$. If a subhalo's $\tau_{merge}$ is below this value it will not be observed at redshift  $z = 0$. Similarly, for the right hand panel we plot the same relation converting the subhalo masses to satellite galaxy masses using the SMHM relation at $z =  1.5$. In both panels the merging time is increased when the ratio of mass between the subhalo/satellite and the central (parent) halo is smaller.

In the next Section \ref{subsec:Frozen} we will first focus on the frozen model. In Section \ref{subsub:EvoInf} we will motivate the need for a finite dynamical time and an evolving SMHM relation, and then in Section \ref{subsec:Tdyn} set constraints on the dynamical times that well reproduce the number and distribution of SDSS satellites as a function of parent host halo mass. In Section \ref{subsec:EvoPro} we will probe the impact of SFR, stripping and quenching in the predicted satellite distribution in the local Universe.

\subsection{Frozen Model}
\label{subsec:Frozen}
We begin by showing the model predictions of satellite number densities and distributions in parent halo mass in a frozen model, where satellites do not evolve in stellar mass after infall.

\subsubsection{Redshift Evolution and Mergers}
\label{subsub:EvoInf}
\begin{figure}
	\centering
	\includegraphics[width = \linewidth]{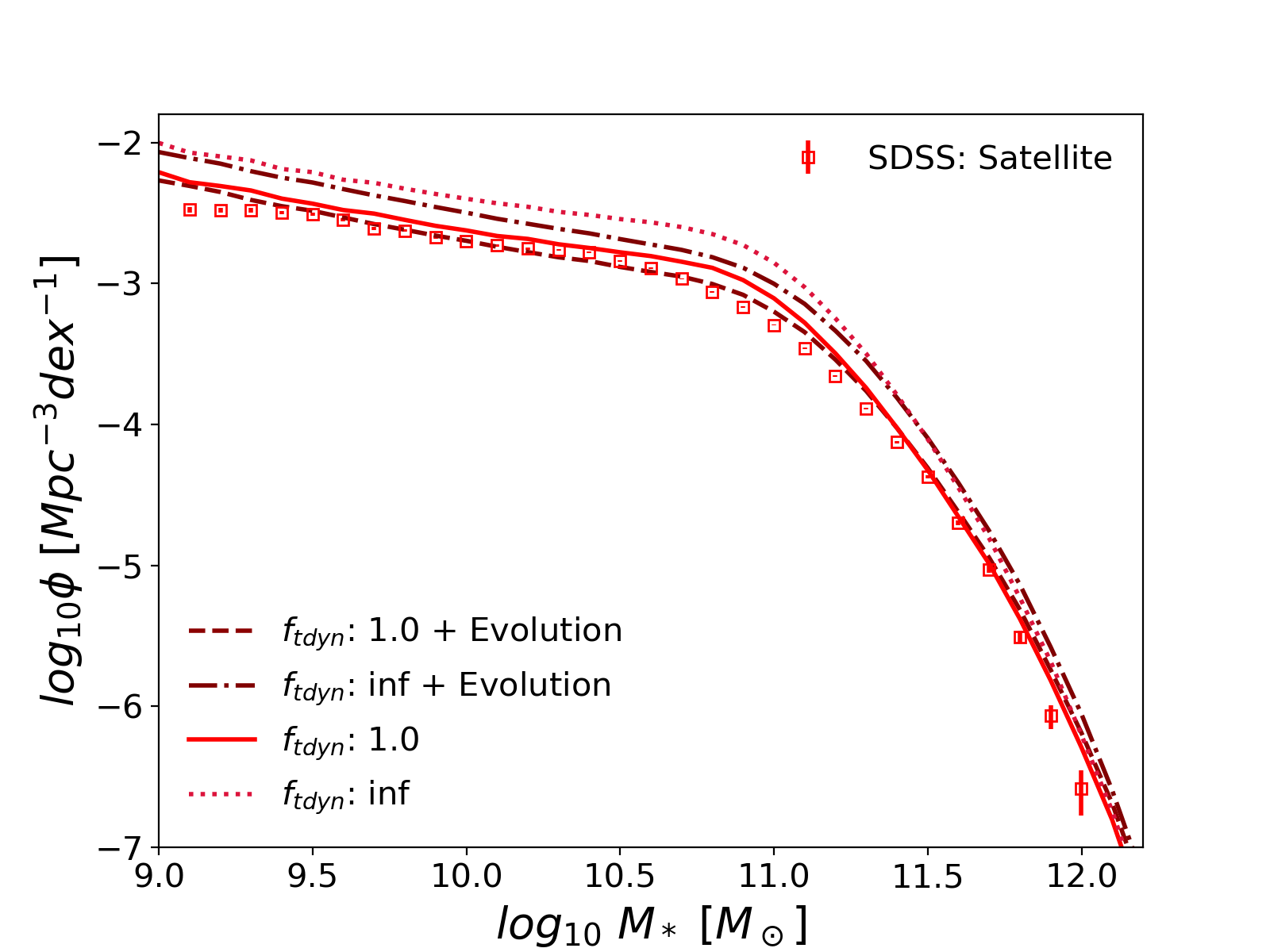}
	\caption{Satellite stellar mass functions generated by the model compared to SDSS data (open squares). The dashed and solid lines refer to models with a reference $f_{tdyn} = 1.0$ (as suggested by N-body simulations), respectively with and without redshift evolution in the input stellar mass-halo mass relation. The dot dashed and dotted lines have infinite $f_{tdyn}$ (i.e. no mergers) with and without redshift evolution in the input stellar mass-halo mass relation.}
	\label{fig:SMF_zEvo}
\end{figure}

We start by exploring a basic model where $\tau_{dyn}$ is infinitely long (no satellite mergers) and the SMHM relation does not evolve in cosmic time, i.e. keeping the same $z=0$ parameters at all redshifts. We choose to start with the simplest assumption of an unevolving SMHM, as also suggested by some empirical evidence at least up to $z\sim 1$ \citep[e.g.,][]{Shankar2014ON1,Buchan2016,Shankar2006NewFormation, Moster2013, Tinker2017TheUniverse,Guo2018EvolutionGalaxies}.
We then compare the outputs of this model with those from models in which we either allow for galaxy-satellite mergers ($f_{t_{dyn}}$ is finite), and/or the SMHM relation evolves with redshift. Figure \ref{fig:SMF_zEvo} shows how the four combinations of these assumptions affect the satellite SMF. For both the models characterized by an infinite dynamical time ($f_{tdyn} = \infty$, dot-dashed and dotted lines), the number densities of satellites are too high compared to the observed satellite SMF (red squares). When setting $f_{tdyn} = 1.0$, i.e. the merger timescales are exactly those parameterized by \citet{McCavana2012TheMergers}, both the evolving and unevolving SMHM models produce very similar number densities and well consistent with the observed satellite SMF. The model with an evolving SMHM relation is particularly successful.

\begin{figure*}
    \centering
    \includegraphics[width = \linewidth]{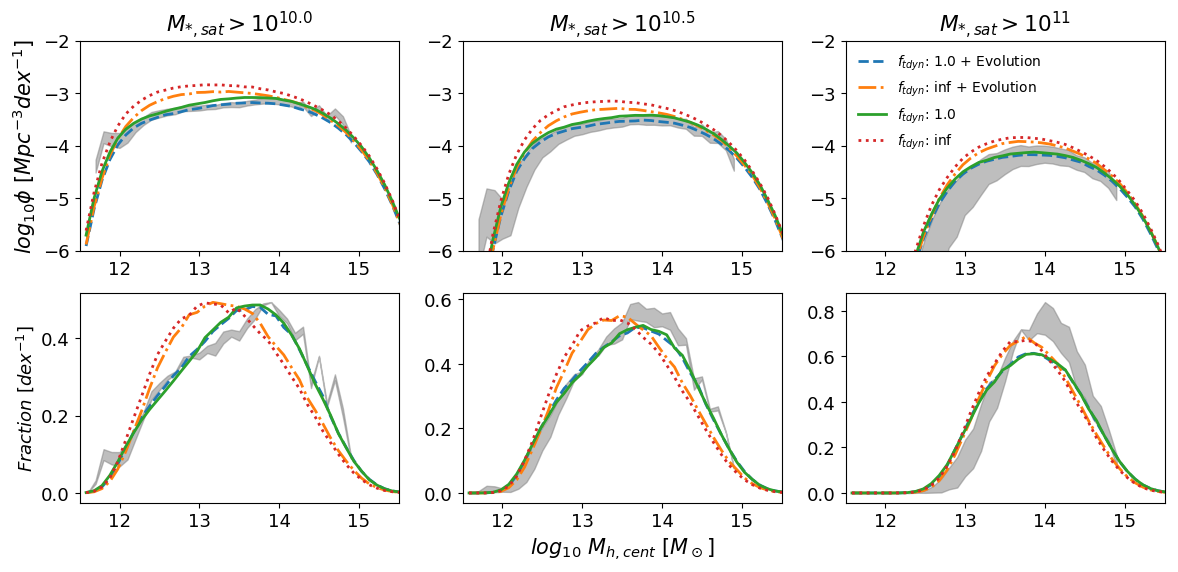}
    \caption{Satellite distributions in parent haloes generated from the model are compared to those observed in SDSS (grey band). Columns from left to right show increasing satellite stellar mass cuts as labelled. The top row shows the number density of satellites expected to be found in each parent halo mass. The bottom row shows the fractional distribution described by Equation \ref{eqn:FracPlot}. The solid and dashed lines show a reference $f_{tdyn} = 1.0$ with and without evolution in the SMHM relation respectively. The dot dashed and dotted lines show $f_{tdyn} = \infty$ (i.e. no mergers) with and without evolution in the SMHM relation respectively. The width of the grey band corresponds to a 10\% uncertainty in satellite stellar masses.}
    \label{fig:Distrbution_zEvo}
\end{figure*}

We now test in Figure \ref{fig:Distrbution_zEvo} the four models against the SDSS distributions of the satellites in groups and clusters. The top row shows the comoving number density per halo mass bin. The bottom row shows the fractional distributions (Equation \ref{eqn:FracPlot}) which better highlights the differences induced by varying model inputs. In both rows the SDSS data are shown as grey bands. As expected, the largest difference is seen between the models with $f_{tdyn} = \infty$ and $f_{tdyn} = 1.0$. When $\tau_{dyn}$ is infinite\footnote{In Python in this case we set this variable using ``numpy.inf''.}, the distribution of satellites skews such that high mass satellites are found more frequently in relatively lower mass parent haloes. We explain the origin of this effect when discussing dynamical time in more detail in Section \ref{subsec:Tdyn}. The evolution\footnote{We direct readers to Appendix \ref{app:Abn} for further discussion and motivation for the redshift evolution parameters used in this work.} of the SMHM relation has a low impact on the number densities. This is in part due to the evolution of the SMHM relation being weak especially in the high mass end, and in part due to most (>60-80\% $z = 0.1$) satellites being accreted at relatively low redshift ($z<1.0$).  We thus support the view that local satellite distributions are largely independent of the specific input of the evolution parameters of the SMHM relation.

\subsubsection{Dynamical Time}
\label{subsec:Tdyn}

\begin{figure}
	\centering
	\includegraphics[width = \linewidth]{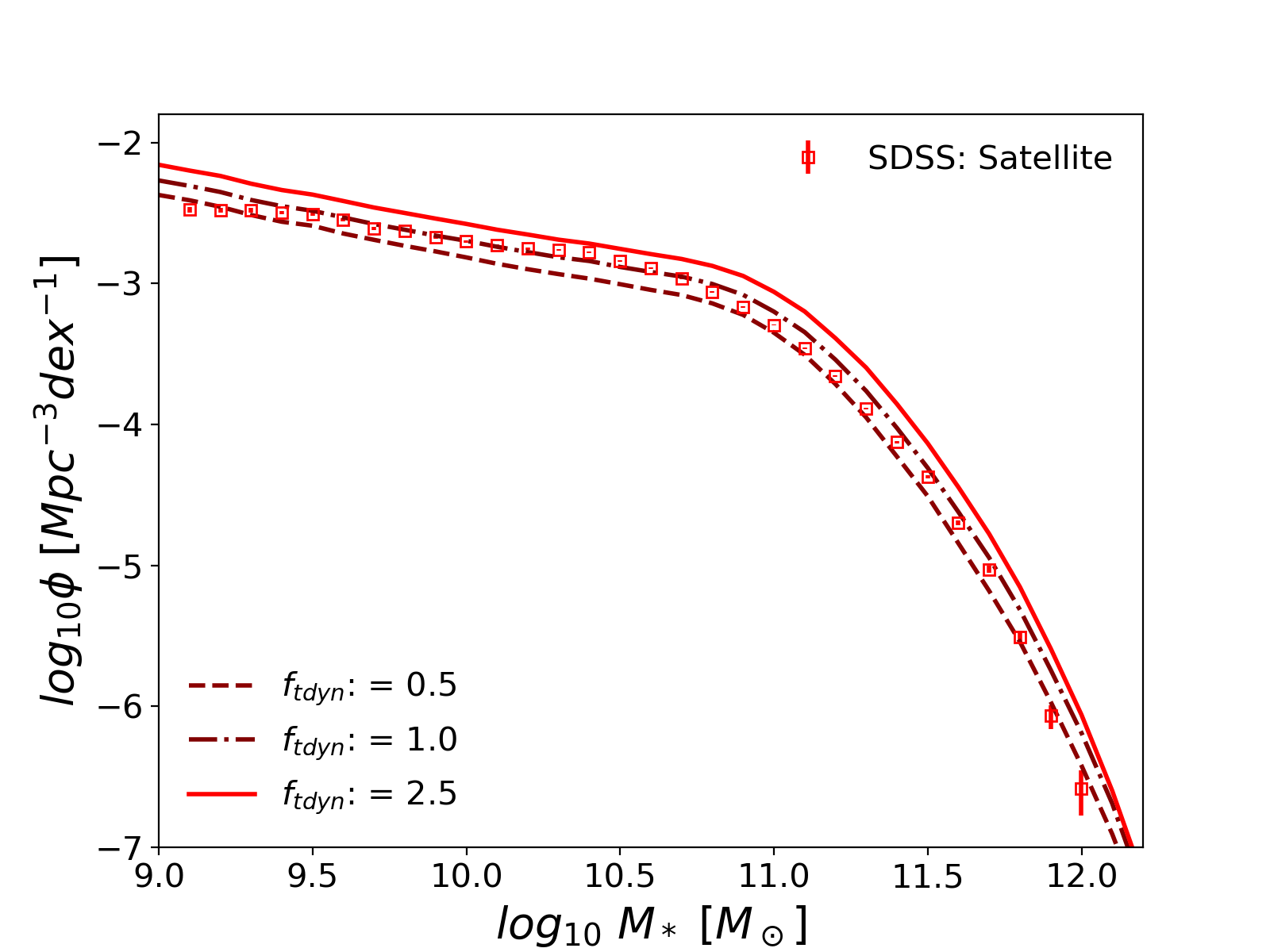}
	\caption{Satellite stellar mass functions generated by the model compared to SDSS data (open squares). The solid, dot dashed, and dashed lines show $f_{tdyn} = 0.5, 1.0,$ and $2.5$ respectively.}
	\label{fig:SMF_Tdyn}
\end{figure}

In this Section we show the predictions of models with different merging timescales, $f_{tdyn}$ = $0.5, 1.0, 2.5$, to probe the effects of dynamical time on the satellite population. In Figure \ref{fig:SMF_Tdyn} we find, as expected, that longer dynamical times tend to increase satellite number densities especially towards lower stellar masses. The number densities of higher-mass satellites are more resilient to increase with dynamical time. In fact, when $f_{tdyn}\gtrsim 1$ the number densities of massive satellites become already very close to the theoretical maximum number density (Figure \ref{fig:SMF_zEvo}) set by $f_{t_{dyn}} = \infty$. It follows that massive satellites are on average a recently accreted population. In other words, there are only a few high mass satellites that had enough time to merge when $f_{tdyn}\gtrsim 1$. In contrast, lower mass satellites have not yet reached their theoretical limit, and thus they still have room to increase their number densities with increasing dynamical time.

\begin{figure*}
	\centering
	\includegraphics[width = \linewidth]{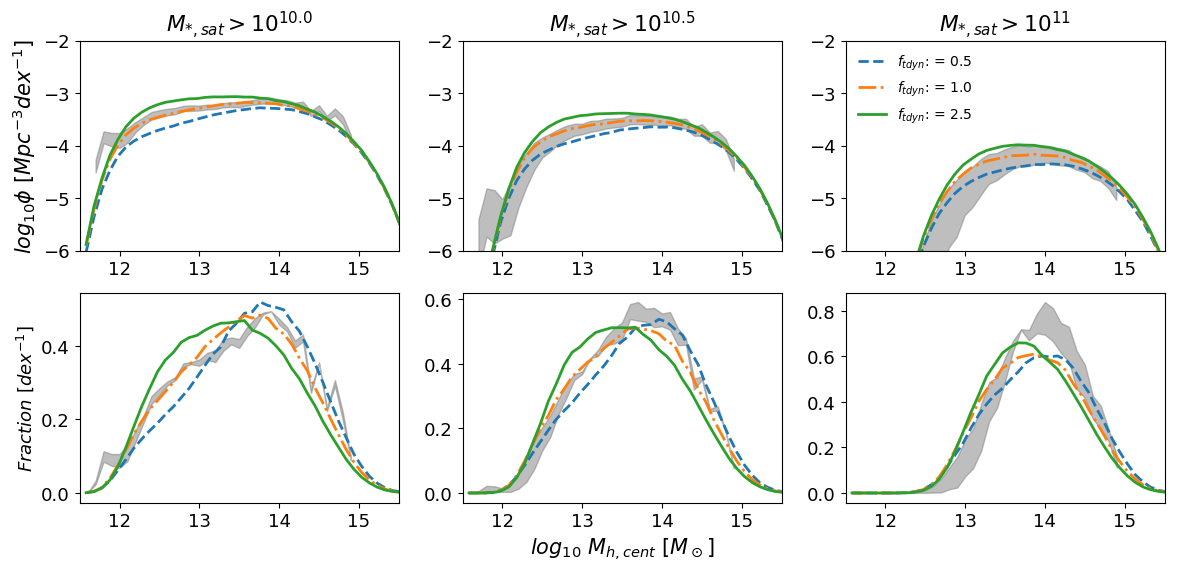}
	\caption{Satellite distributions in parent haloes generated from the model are compared to those observed in SDSS (grey band). Columns from left to right show increasing satellite stellar mass cuts as labelled. The top row shows the number density of satellites expected to be found in each parent halo mass. The bottom row shows the fractional distribution described by Equation \ref{eqn:FracPlot}. The dashed, dot dashed, and solid lines show $f_{tdyn} = 0.5, 1.0,$ and $2.5$ respectively. The width of the grey band corresponds to a 10\% uncertainty in satellite stellar masses.}
	\label{fig:Distrbution_Tdyn}
\end{figure*}

Similarly to Figure \ref{fig:Distrbution_zEvo}, Figure \ref{fig:Distrbution_Tdyn} shows how the distribution of satellites as a function of parent halo mass is affected by $f_{tdyn}$. The number density distribution (top row) shows results similar to the satellite SMF where increasing dynamical time increases the number densities. However, there is also an apparent steepening effect for which lower mass host haloes end up containing relatively less satellites with respect to models with longer dynamical times. The fractional plot (bottom row) accentuates this change in the number density distributions shown in the top row: shorter dynamical times shift the peak of the distribution to the right as relatively more satellites are observed in high mass host haloes.

This steepening of the satellite distribution as a function of halo mass, as well as the shift mentioned in Section \ref{subsub:EvoInf} where infinite dynamical times move satellites preferentially to lower halo masses, are both caused by the amount of time satellites survive in their hosts. Massive satellites are far more common in massive hosts, as can be inferred from the SHMF. Therefore, irrespective of the chosen merging timescale, there will always be a high number density of surviving massive satellites in higher mass parent haloes. However, when merging timescales are increased, the lower number densities of massive satellites in moderately-sized haloes are also increased. Given that lower mass parent haloes are more abundant, the reduction of merging timescales tends to shift the peak of the fractional distribution of galaxies to lower mass parents. Otherwise put, the merging timescales in clusters ($ \log_{10} M_{h,cent}$ $[M_{\odot}]$ $> 13.75$) are so long that even a factor five reduction in merger timescale still does not give the satellite galaxies sufficient time to merge with the central galaxy. This effect can be inferred from Figure \ref{fig:Tdyn_M}, where we see steeper gradients for a given subhalo/satellite mass with increasing parent halo mass. Merging is more efficient for the lower mass satellites as can be seen by the steepness of the $f_{t_{dyn}} = 0.5$ model (dashed lines) in the top row of Figure \ref{fig:Distrbution_Tdyn}.

The least-square residuals to the SMF, number density distribution and fractional distributions are given in Table \ref{tab:bestfit}. There is no model that simultaneously fully matches all the observations in all mass ranges. The trend in both the number density distribution and the fractional distribution is that slightly longer dynamical times ($f_{tdyn}  = 1.2$) are favored by the less massive satellites. Longer dynamical timescales better match the halo mass distributions (number density distribution/fractional distributions) for lower mass satellites, and vice versa for higher mass satellites with $\log M_{*}/M_{\odot} >11$. Nevertheless, the simple combination of abundance matching and dynamical merging timescales as suggested by pure N-body simulations ($f_{t_{dyn}} = 1.0$) tends to provide overall good agreement to both the satellite SMF and the satellite distributions, without the need to invoke additional physics in the (late) evolution of satellite galaxies after infall.

\begin{table*}
\centering
\caption{We show the sum of the squared residuals between the SDSS and our model. The satellite SMF is calculated between 9.1 and 12.0 $M_{*}$. The SDF fit is calculated between 12 and 14.9 $M_h$ for the >10 and >10.5 plots, and between 12.5 and 14.9 $M_h$ for >11. The Fractional plot fit is calculated between 11.6 and 14.9 $M_h$.}
\label{tab:bestfit}
\begin{tabular}{c|c|ccc|ccc}
$f_{t_{dyn}}$   & SSMF   (Fig \ref{fig:SMF_Tdyn})               & \multicolumn{3}{c}{SDF  } \vline & \multicolumn{3}{c}{Fractional Distribution } \\
   &   (Fig \ref{fig:SMF_Tdyn})               & \multicolumn{3}{c}{ (Top Row Fig \ref{fig:Distrbution_Tdyn}) } \vline & \multicolumn{3}{c}{ (Bottom Row Fig \ref{fig:Distrbution_Tdyn})} \\ \hline
            \multicolumn{1}{l}{} \vline & \multicolumn{1}{l}{} \vline & \multicolumn{1}{l}{\textgreater{}10} & \multicolumn{1}{l}{\textgreater{}10.5} & \multicolumn{1}{l}{\textgreater{}11} \vline & \multicolumn{1}{l}{\textgreater{}10} & \multicolumn{1}{l}{\textgreater{}10.5} & \multicolumn{1}{l}{\textgreater{}11} \\ \hline
0.5    & 0.022   & 0.19  & 0.55    & 0.073    & 0.0042 & 0.0047 & 0.0078  \\
0.8    & 0.025   & 0.13  & 0.51    & 0.089    & 0.0020 & 0.0017 & 0.0054  \\
1.0    & 0.034   & 0.12  & 0.56    & 0.10     & 0.0015 & 0.0011 & 0.0050  \\
1.2    & 0.043   & 0.12  & 0.53    & 0.11     & 0.0015 & 0.00094& 0.0046  \\
1.5    & 0.054   & 0.12  & 0.52    & 0.13     & 0.0017 & 0.0010 & 0.0045                              
\end{tabular}
\end{table*}

\subsection{``Non-Frozen'' evolutionary models}
\label{subsec:EvoPro}
We showed in the Section \ref{subsec:Frozen} that the number and distribution of satellite galaxies are primarily driven by the accretion history and dynamical friction timescale, and depend weakly on the exact redshift evolution in the input SMHM relation. 

However, as discussed in Section \ref{sec:SatEvo}, satellites may not be a strictly frozen population. After infall a number of processes such as star formation, quenching and/or stellar stripping, may all well affect the final stellar masses of satellite galaxies. In this Section we explore the impact that each one of these processes could have on the predicted local population of satellite galaxies. We will prove, in particular, that additional, late stellar stripping, star-formation and quenching, whilst not necessarily playing a major role in varying integrated number densities, are still all necessary ingredients to fully reproduce SDSS data in terms of, for example, specific star formation rate distributions.

In what follows we explore the two star formation rate models with the quenching model described in Sections \ref{sec:SFR} and \ref{sec:QuenchingModels}. The first, purely observationally-based, star formation rate model strictly follows the SFR parametrization by \citet[][T16 hereafter]{Tomczak2016THE4}, given in Equation \ref{eqn:SFR}. The second star formation rate model (which we label as ``CE'') is instead based on a continuity equation approach similar to \cite{Leja2015RECONCILINGFUNCTION}. In essence, the latter is based on first assuming number conservation of galaxies, and then deriving the implied star formation rates as a function of stellar mass from the time growth in any stellar mass bin implied by the redshift evolution of the stellar mass function (corrected for gas loss from dying stars assumed to be an average of 40\%). The novelty in the latter model with respect to previous work is that we do not tune the resulting star formation rate on the total but rather only on the stellar mass function of \textit{central} galaxies, which is in turn iteratively constrained by matching the local stellar mass function of SDSS centrals. This has the main effect of lowering the implied average star formation rate at fixed stellar mass at any given epoch (see Equations \ref{eqn:SFR} and \ref{eqn:SFR_CE}).

\begin{figure}
	\centering
	\includegraphics[width = \linewidth]{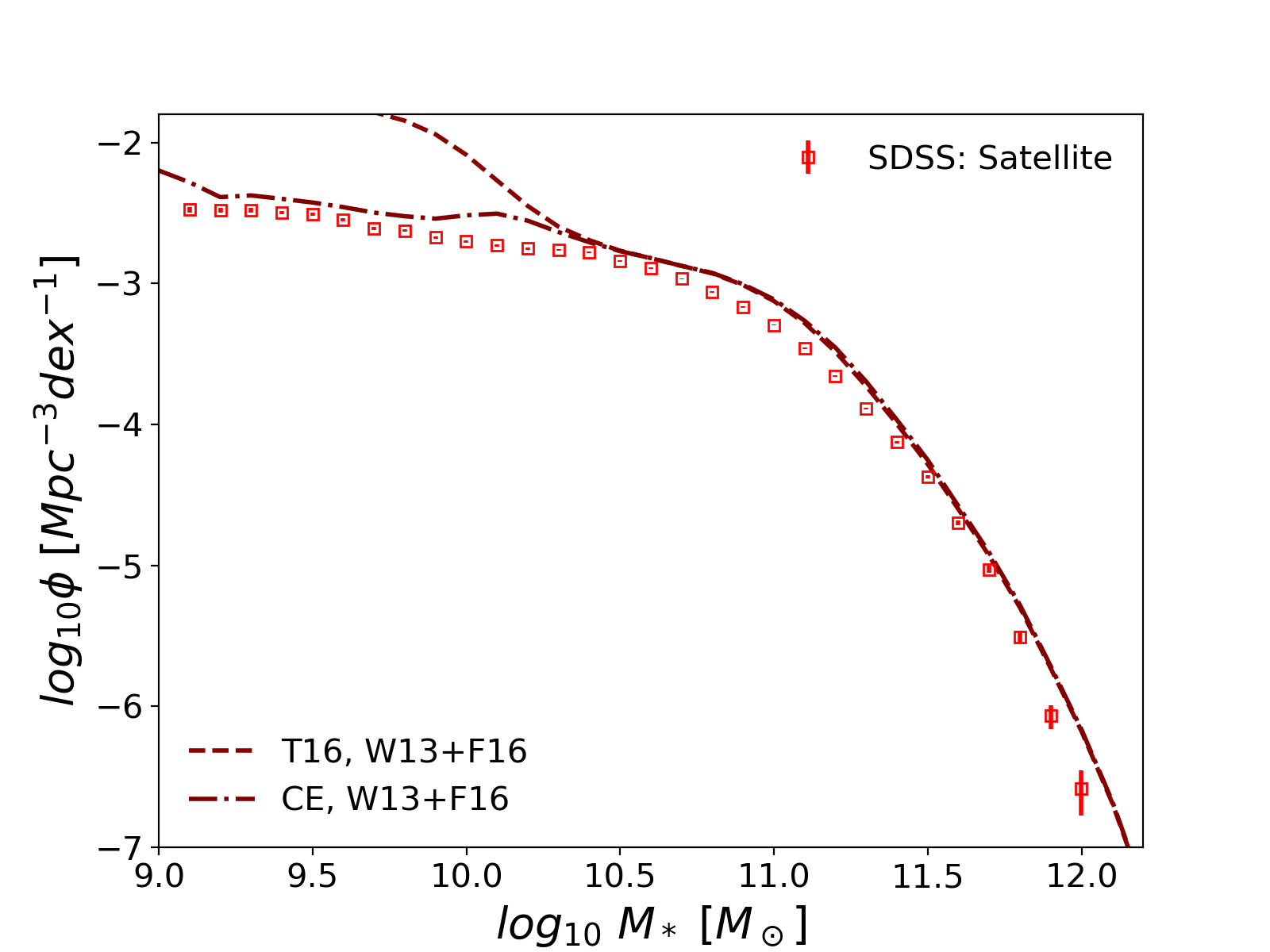}
	\caption{Satellite stellar mass functions generated from the model using both the \citet{Tomczak2016THE4} (dashed) and continuity (dot dashed) star formation rates compared to the SDSS satellite stellar mass function (open squares).}
	\label{fig:SMF_SF_Q}
\end{figure}

We compare the satellite SMFs produced by the two star formation+quenching models addressed above to our SDSS stellar mass function of satellites in Figure \ref{fig:SMF_SF_Q}. It is apparent that 
using the observed SFR by T16 (dashed line), even inclusive of the best recipes for quenching, still substantially overproduces the number density of galaxies below $M_* \lesssim 3\times 10^{10}\, M_{\odot}$. This is a well-known problem affecting the full (dominated by central) galaxy population \citep[e.g.,][]{Leja2015RECONCILINGFUNCTION}: the integrated (observed) SFR is not consistent with the moderate growth over time of the SMF causing an overproduction of galaxies becoming gradually more severe at lower stellar masses. Our results point to a similar problem affecting the satellite population, on the assumption that the latter at infall share the same SFR distribution as a typical central galaxy of the same stellar mass.

\begin{figure*}
	\centering
	\includegraphics[width = \linewidth]{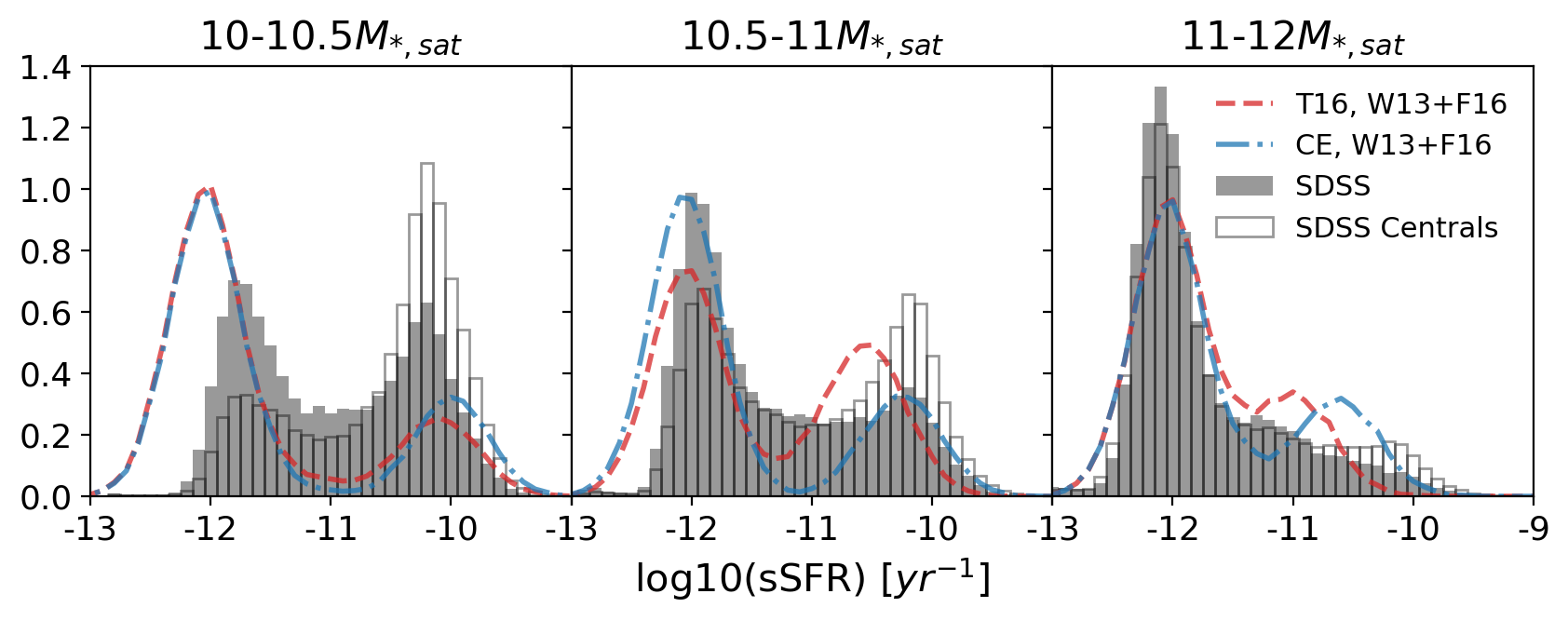}
	\caption{Specific star formation rate distributions in three mass ranges (as labelled) generated from the model using both the \citet{Tomczak2016THE4} (dashed) and continuity (dot dashed) star formation rates. Data shown are the distribution of specific star formation rate in SDSS satellites (shaded bars) and SDSS centrals (unfilled bars).}
	\label{fig:SSFR}
\end{figure*}

To further constrain our adopted continuity equation-based star formation rate distributions, we compare our best models with the bimodality in (satellite) SSFR measured in the SDSS data (grey shading in Figure \ref{fig:SSFR}). The data show a clear bimodality in SSFR at least for galaxies below $M_*\lesssim 10^{11}\, M_{\odot}$, at variance with the SSFR distribution of centrals (black thick long dashed lines), which point to a much larger population of starforming galaxies. This difference in SSFR distributions has been often ascribed to environmental quenching \citep[e.g.][]{Peng2010MassFunction}, either in the form of stripping or strangulation. Both the star formation rate reference models predict very similar distributions in SSFR for galaxies above $M_*\gtrsim 3\times10^{10}\, M_{\odot}$, as expected from the invariance in the predicted number densities above this stellar mass. When using the continuity equation-based method (CE, dot dashed line) the overproduction is greatly reduced and the SSMF is very close to the SSMF with no evolution in the satellite population.

\begin{figure}
	\centering
	\includegraphics[width = \linewidth]{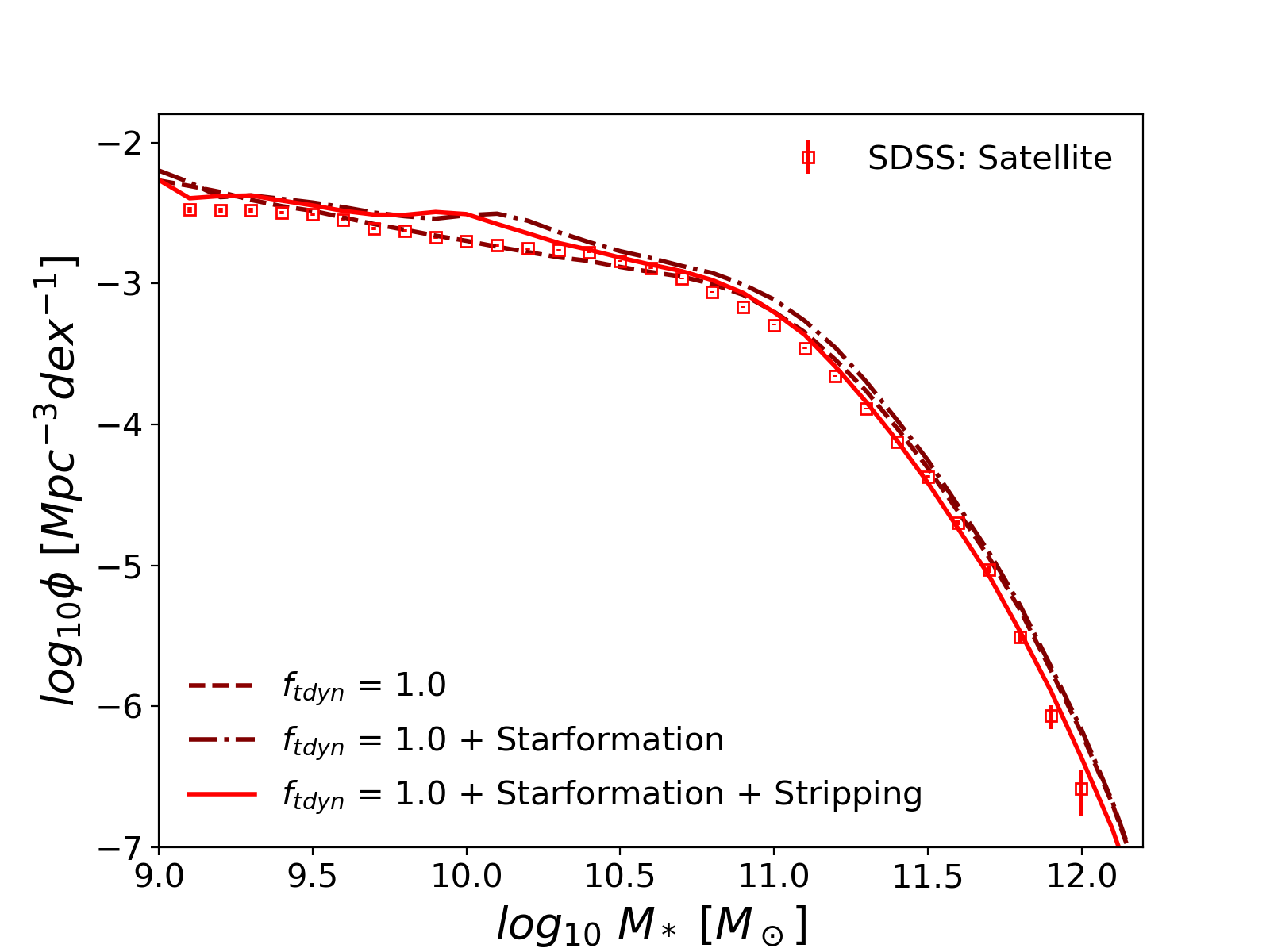}
	\caption{Satellite stellar mass functions generated from the model compared to SDSS satellites (open squares). The models shown all have $f_{tdyn} = 1.0$ and are the reference 'frozen model' (dashed line), starformation (CE model) only (dot dashed line) and starformation and stripping (solid line).}
	\label{fig:SMF_SF_Strip}
\end{figure}

\begin{figure*}
	\centering
	\includegraphics[width = \linewidth]{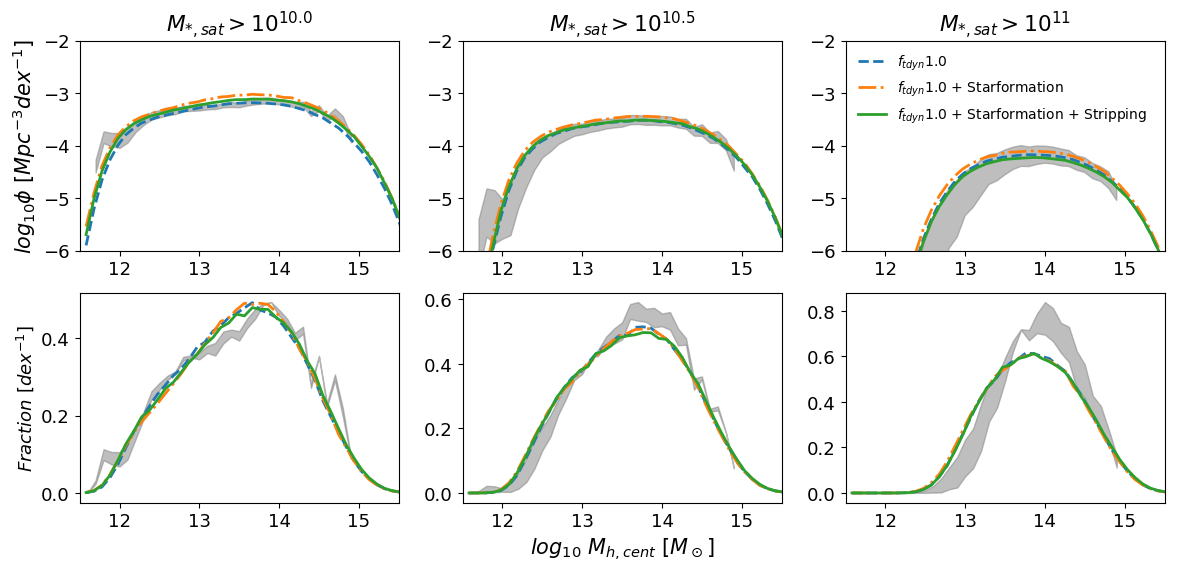}
	\caption{Satellite distributions in parent haloes generated from the model are compared to those observed in SDSS (grey band). Columns from left to right show increasing satellite stellar mass cuts as labelled. The top row shows the number density of satellites expected to be found in each parent halo mass. The bottom row shows the fractional distribution described by Equation \ref{eqn:FracPlot}. The models shown all have $f_{tdyn} = 1.0$ and are the reference 'frozen model' (dashed line), starformation only (dot dashed line) and starformation and stripping (solid line). The width of the grey band corresponds to a 10\% uncertainty in satellite stellar masses.}
	\label{fig:Distribution_SF_Strip}
\end{figure*}

We now move on showing the relative impact of star formation rate and stellar stripping on the satellite stellar mass function. Figure \ref{fig:SMF_SF_Strip} and Figure \ref{fig:Distribution_SF_Strip} show stellar mass function and host halo mass distributions for the $f_{tdyn} = 1.0$ reference model with no stripping nor star-formation, the CE star formation model, and the CE star formation model with stripping (long-dashed, dot-dashed, and solid lines, respectively). We see from both Figures that the reference and star-formation model are almost indistinguishable. The stripping, at least at the level implemented in this work, also has a rather minor effect, at the most reducing the number densities of the most massive satellites ($>10^{11}M_{\odot}$) by $\lesssim 0.2$dex. Note that that the \textit{cumulative} effect of stripping on the stellar mass growth of central galaxies could be much more prominent, as emphasized by a number of groups \citep[e.g.,][]{Cattaneo2017TheSimultaneously}.

\begin{table*}
\centering
\caption{We show the sum of the squared residuals between the SDSS and our model as in \ref{tab:bestfit} with the same mass ranges for the fitting. All models have $f_{t_{dyn}} = 1.0$ from top to bottom we then have the reference frozen model, the model with starformation, and the model with stripping and starformation.}
\label{tab:SF_Strip}
\begin{tabular}{c|c|ccc|ccc}
$f_{t_{dyn}}$   & SSMF   & \multicolumn{3}{c}{SDF  } \vline & \multicolumn{3}{c}{Fractional Distribution } \\
   &   (Fig \ref{fig:SMF_SF_Strip})               & \multicolumn{3}{c}{ (Top Row Fig \ref{fig:Distribution_SF_Strip}) } \vline & \multicolumn{3}{c}{ (Bottom Row Fig \ref{fig:Distribution_SF_Strip})} \\ \hline
            \multicolumn{1}{l}{} \vline & \multicolumn{1}{l}{} \vline & \multicolumn{1}{l}{\textgreater{}10} & \multicolumn{1}{l}{\textgreater{}10.5} & \multicolumn{1}{l}{\textgreater{}11} \vline & \multicolumn{1}{l}{\textgreater{}10} & \multicolumn{1}{l}{\textgreater{}10.5} & \multicolumn{1}{l}{\textgreater{}11} \\ \hline
1.0 & 
0.034 & 0.12 & 0.53 & 0.10 & 0.0015& 0.0011& 0.0049\\
\begin{tabular}[c]{@{}c@{}}1.0\\ With Star Formation\end{tabular} & 
0.049 & 0.077& 0.43 & 0.14 & 0.0018& 0.0012& 0.0059\\
\begin{tabular}[c]{@{}c@{}}1.0\\ With Stripping and Star Formation\end{tabular} & 
0.021 & 0.087& 0.47 & 0.088& 0.0016& 0.0015& 0.0056
\end{tabular}
\end{table*} 

Table \ref{tab:SF_Strip} shows the sum of the square residuals to the SMF, SDF and fractional distributions for the same models discussed above, our reference frozen one, and the one with evolution of satellites after infall (stellar stripping and continuity equation-based star formation). Table \ref{tab:SF_Strip} shows that the satellite late evolution has little effect on the fractional distribution. In the number density distribution we see an improved fit for galaxies in the $M_*>10.^{11}\, M_{\odot}$ range, mainly induced by the stripping which slightly reduces the number density of massive galaxies. Table \ref{tab:SF_Strip} shows the sum of square residuals for the dynamical time with $f_{t_{dyn}}=1$, for the frozen and evolved models. 

In summary, our results point to the fact that direct abundance matching between centrals and satellite galaxies and the halo plus subhalo mass functions in the local Universe (and at higher redshifts) is a good approximation to the mean stellar mass-halo mass relation for central galaxies, irrespective of the precise late evolution of satellites after infall.

\section{Discussion}
\label{sec:discu}

\subsection{Strengths of the statistical approach with respect to the state of the art}
\label{subsec:discu_strenghts}

The statistical semi-empirical model introduced in this work presents a highly complementary approach to the existing suit of cosmological galaxy evolution models. The latter models, either being fully analytic or fully numerical, inevitably rely on large boxes to simulate both low and high mass galaxies with sufficient statistics. Even traditional semi-empirical models, based on applying abundance matching techniques to vast catalogues of dark matter merger trees \citep[e.g.,][and references therein]{Behroozi2018UniverseMachine:Z=0-10}, still require large computational resources. Our approach, based on statistical mean dark matter accretion histories, allows to explore the full range of galaxy stellar and halo masses without the need to simulate large volumes or even weight dark matter merger trees. Whilst we lose the ability to track an individual galaxy through cosmic time, we are able to rapidly predict the statistics of any subpopulation of chosen galaxies at any redshift and environment. 

There is vast literature on the modelling of satellite galaxies. Here we recall just a couple of examples of semi-empirical and semi-analytic models to highlight some of the key similarities and key differences. \citet{Neistein2013A2011}, with an approach similar to ours, separated the central and satellite populations in an attempt to better define the galaxy halo connection. By allowing in an N-body simulation the stellar mass of satellite halaxies to depend on both the host subhalo mass and on the parent halo mass, \citet{Neistein2013A2011} find that the local satellite stellar mass-halo mass is substantially less well defined than the one for central galaxies. In our model satellites insted strictly follow the stellar mass-halo mass relation of centrals at infall. In this way we find the resulting satellite distributions to be well reproduced. Our semi-empirical statistical model was able to reproduce multiple observables such as the stellar and parent halo mass distributions, with essentially only one parameter, $f_{t_{dyn}}$. By working with minimal assumptions and related free parameters, our approach is thus less prone to possible degeneracies affecting more traditional, multi-parameter techniques.

Another key difference with respect to previous models concerns ``orphan galaxies''. In N-body or merger tree based simulations, when a subhalo goes below the resolution limit, an orphan galaxy is created \citep[e.g.,][]{Guo2011FromCosmology, DeLucia2011TimesCosmology}. It is then necessary to make an assumption on how much longer that subhalo (and hosted satellite galaxy) will survive. In our model we avoid this complication by self-consistently assigning to all satellites a (full) observability timescale at infall.

\subsection{Future applications of the statistical semi-empirical model}
\label{subsec:discu_future}

We have shown here the ability of the statistical semi-empirical model to reproduce the observed group and cluster richness at $z=0.1$ using simply the dynamical friction timescale from N-body simulations. In future work we will also compare our model outputs with the richness of groups and clusters at high redshift thus providing additional important constraints to models of structure formation. For example, we will use our statistical modelling to fit extreme objects such as the high redshift ($z=2.5$) massive cluster reported in \citet{Wang2016DISCOVERYZ=2.506}, which has been claimed to be challenging in other $\Lambda CDM$ cosmology-based models. By predicting the (correct) distributions of satellite galaxies predicted over many epochs, more accurate merger rates and pair fractions can be calculated and compared to observations found in surveys such as those reported in \citet{Mundy2017A3.5}. Via more robust estimates of galaxy merger rate we will be able to set upper limits on the impact of mergers in shaping galaxy morphology \citep{Hopkins2009TheDemographics}, build galactic bulges \citep{Cole2000,Hopkins2010MERGERSMATTER,Shankar2013SizeUniverse} and trigger AGN activity \citep{Villforth2017HostLuminosities}.

Our technique being fast and flexible is an extremely valuable resource in view of the next large and deep extra-galactic surveys for example Euclid \citep{Refregier2010EuclidBook}. Our model can be both predictive, working out specific expectations for such surveys, and then reactive, using the model outputs as constraints on existing formation models.

\section{Conclusions}
\label{sec:conclu}

In this paper we have presented 'STEEL' a \textit{STatistical} sEmi-Empirical modeL that replaces the traditional merger tree dark matter backbone with a ``statistical accretion history''. In essence, we first trace backwards the mean accretion history of dark matter haloes in a given bin of halo mass at the redshift of interest $z$. We then gradually build in time their satellite population by integration of the subhalo mass function, and at each time step assign stellar masses to subhaloes via abundance matching techniques. Our approach is extremely fast and accurate, allowing to probe the galaxy number densities and average properties within the full range of stellar masses probed by large extragalactic surveys, with virtually no limitation on volume size or mass resolution. 

We use the statistical semi-empirical model to predict the satellite galaxy population at stellar masses above $M_*\gtrsim 10^{9}\, M_{\odot}$. We find that irrespective of the exact input stellar mass-halo mass relation, the main driver shaping the local stellar mass function of satellite galaxies is by far the input dynamical friction timescale. In particular, adopting a merging timescale very close to the one calibrated from high-resolution N-body dark matter simulations, provides an excellent match to the satellites stellar mass function as inferred from SDSS. Shorter dynamical friction timescales not only reduce the number densities of satellites (as more satellites disappear merging with central galaxies), but also shift the abundances of satellites into higher mass parent haloes than actually observed. Conversely, longer dynamical friction timescales create an overproduction in the satellite stellar mass function and induce too many satellites in lower mass parents. Thus a traditional hierarchical dark matter cosmology naturally provides the right abundances of satellites in the local Universe without any additional fine-tuning.

We then incorporate popular recipes for star formation, stellar stripping and quenching in satellite galaxies after infall. We find that, within reasonable variations of the input parameters, all these processes play a secondary role. At fixed merging timescale the resulting stellar mass function and halo mass distribution of local satellites end up being very similar. All in all, our results highly suggest that at any given epoch, a direct abundance matching between the total galaxy stellar mass function and halo plus subhalo mass functions, provides a very good approximation to the mean stellar mass-halo mass relation of central galaxies, irrespective of the exact (late) evolution of satellites after infall. In other words, our results confirm the common assumption that at any given epoch satellites can be abundance matched as centrals at infall, irrespective of their redshift of infall.

We find that, similarly to what previously noticed for centrals, adopting the latest empirical determinations of the star formation rate as a function of stellar mass and redshift significantly overproduces the abundance of galaxies below $M_* \lesssim 3\times 10^{10}\, M_{\odot}$, irrespective of the quenching, stripping, or merger timescales adopted. A star formation rate as predicted by a continuity equation approach well matches both the local abundances of satellites and the bimodality in specific star formation rates for all galaxies above $M_*\gtrsim 10^{11}\, M_{\odot}$ as measured in the Sloan Digital Sky Survey.

STEEL is an ideal tool to rapidly compute accurate galaxy merger rates as a function of stellar mass, time and environment. It can reveal the impact of mergers in shaping the structural and dynamical evolution of galaxies, and eventually their central supermassive black holes. We will look into some of these topics in more detail in future work.

\section*{Acknowledgements}

We warmly thank B. Moster and J. Leja for detailed discussions on the SMHM relation and the SFR respectively. We also thank C. Conselice, N. Menci and C. Marsden for useful discussions on the project. We also acknowledge extensive use of the Python libraries astropy, matplotlib, numpy, pandas, and scipy. PJG acknowledges support from the STFC for funding this PhD.




\bibliographystyle{mnras}
\bibliography{Mendeley.bib} 



\appendix

\section{Abundance Matching}
\label{app:Abn}
\begin{figure*}
	\centering
	\includegraphics[width = \linewidth]{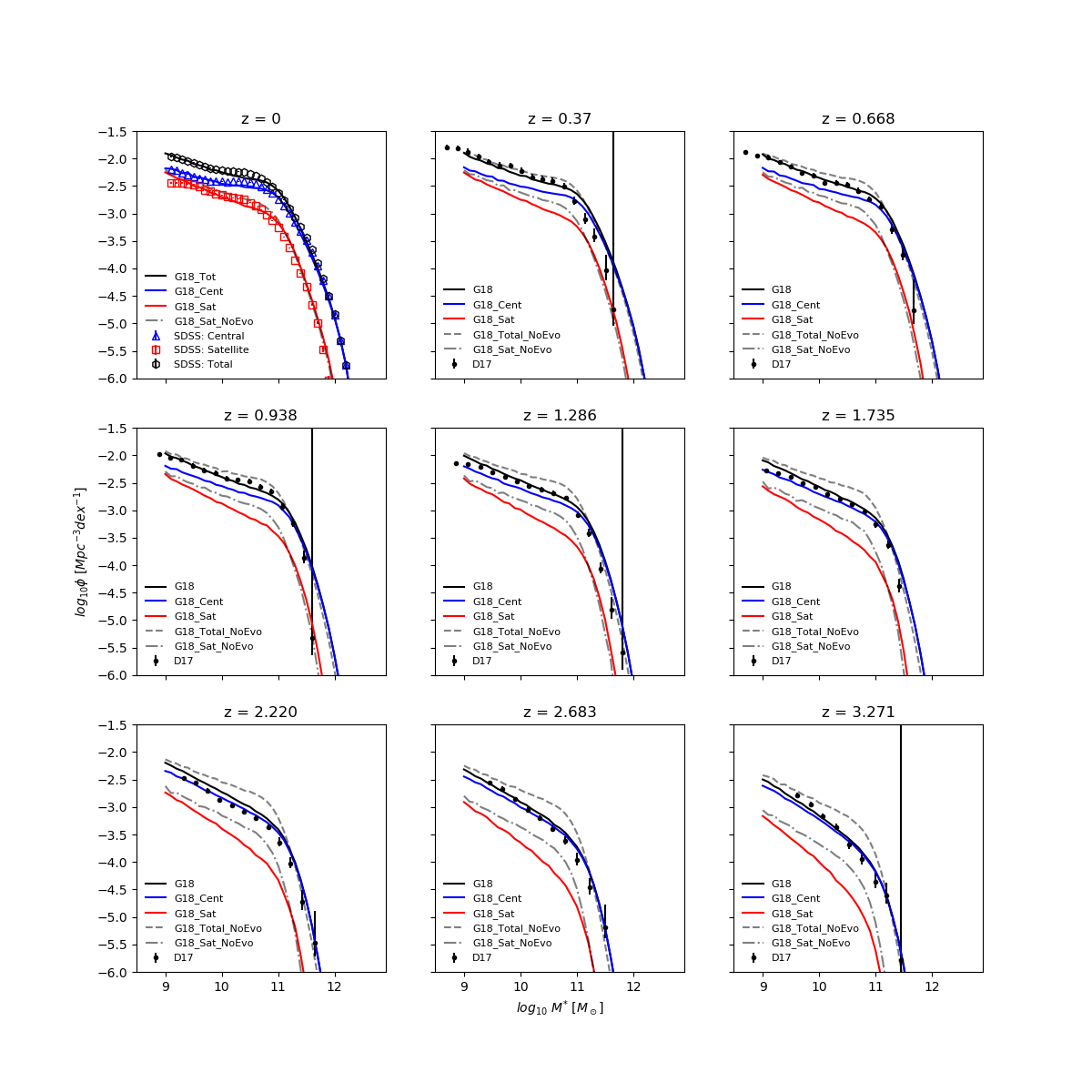}
	\caption{We show the results of the redshift dependent abundance matching at eight different epochs above redshift $z = 0$. The SDSS data is shown in the redshift $z = 0.1$ panel with crosses, higher redshift panels show COSMOS2015 (D17) data as green dots. The final result is shown  in each panel as green blue and red lines for total, central, and satellite galaxies respectively. The no-evolution is shown in gray in all panels. In this plot G18 refers to the central SMHM fit from this work.}
	\label{figA1:SMF_Abn_z}
\end{figure*}

In Figure \ref{figA1:SMF_Abn_z} we show the central and satellite stellar mass functions from the evolving abundance matching relation described in Section \ref{sec:Abn}. We show nine different redshifts, as labelled, using the frozen model with a dynamical-time factor $f_{tdyn} = 1.0$, which best matches the local data from SDSS. The $z>0.3$ data are from \cite{Davidzon2017TheSnapshots}.

A limitation of the present methodology is that it lacks, at present, robust observational estimates of the high-redshift central and satellite stellar mass functions to better anchor our fitting procedure. 
Nevertheless, we see in Figures \ref{fig:Distrbution_zEvo} and \ref{fig:SMF_zEvo}, the satellite SMF, the number density distribution or the fractional distribution do not necessarily require any redshift evolution in the input parameters of the SMHM relation to fit the local data. This is mainly due to the fact that most ($>60\%$) of the accretion of the satellites observed at $z=0$ occurs at relatively recent times, $z<0.5-1$. Redshift evolution is however required for a good fit to the data when $z > 1$, as demonstrated by the grey lines in Figure \ref{figA1:SMF_Abn_z}.

\section{Star Formation Rate From A Continuity Equation Approach}
\label{app:SFRCont}
\begin{figure}
	\centering
	\includegraphics[width = \linewidth]{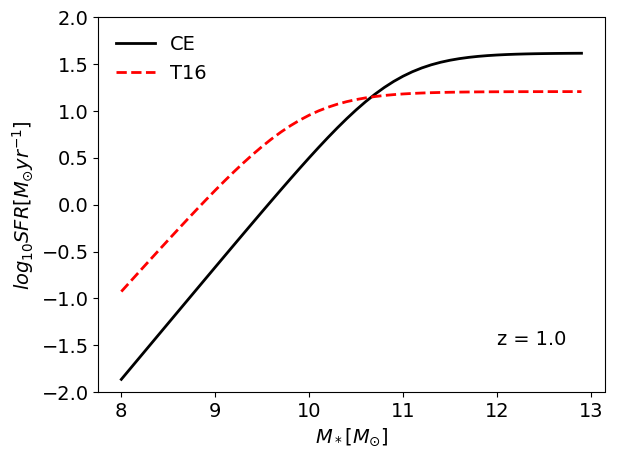}
	\caption{The Tomczak (red dashed) and continuity equation approach (black solid) star formation rate - stellar mass relations as given in Section \ref{sec:SFR} at redshift $z = 1$.}
	\label{figA2:SFR}
\end{figure}

A general star formation rate continuity equation model considers the change in the stellar mass function over time step $\delta t$. Assuming galaxies maintain rank order the model calculates the average star-formation rate a galaxy would have to maintain the implied growth in stellar mass. In this work, using a continuity method, as described above, we generate a SFR that is self consistent with the HMF and abundance matching used in this work. Usually, the input SMF used is an observed quantity. However we use as input for our continuity equation the central SMF generated by the SMHM relation from Section \ref{sec:Abn} and the central HMF from \cite{Tinker2010THETESTS}\footnote{We also note that given the release of resolved centrals at high redshift, improved abundance matching, and/or improved local SMF the SFR (as well as the rest of the model) is simply adjusted self consistently as is the power of a semi-empirical model.}. When considering the mass growth we must consider mass loss, else the SFR is massively under-predicted. We use a instantaneous loss fraction of $40\%$. We also neglect the mass gained from mergers. Mergers would decrease the SFR calculated (and therefore provide a better fit though out this work) so SFR from this method should be considered as an upper limit. We show in Section \ref{sec:SatEvo} our SFR has limited effect on the results of this work. We alter the parameters in Equation \ref{eqn:SFR} to fit our continuity equation-driven $M_* - SFR$ relations and the new fit is given in Equation \ref{eqn:SFR_CE}. During this fit we ignore the SFR inferred from galaxies that sit above the knee of the SMHM relation where the growth becomes merger dominated \citep{Tomczak2016THE4}. We show the difference between the continuity equation approach and the Tomczak SFR in Figure \ref{figA2:SFR}.
\section{Glossary of Technical Acronyms}
\label{app:TechAcro}
Throughout this work we use several technical acronyms. We provide the reader with a glossary of these in Table \ref{tab:Glos} to aid reading of the paper and equations therein.
\begin{table}
\label{tab:Glos}
\caption{A list of technical acronyms used throughout this paper.}
\begin{tabular}{ll}
Acronym       & Definition                           \\ \hline
STEEL         & The name of our model: STatistical sEmi-Empirical modeL \\
HMF           & Halo Mass Function                         \\
SHMF          & Sub-Halo Mass Function                     \\
USHMF         & Unevolved Sub-Halo Mass Function           \\
USSHMF        & Unevolved Surviving Sub-Halo Mass Function \\
$\delta$USHMF & Unevolved Sub-Halo Mass Function Accretion \\
SMF           & Stellar Mass Function                      \\
SSMF          & Satellite Stellar Mass Function            \\
SFR           & Star Formation Rate                        \\
SFH			  & Star Formation History					   \\
MLR			  & Mass Loss Rate
\end{tabular}
\end{table}


\bsp	
\label{lastpage}
\end{document}